\documentclass[11pt,a4paper,DIV=11,numbers=noenddot]{scrartcl}
\usepackage[utf8]{inputenc}
\makeatletter
\DeclareOldFontCommand{\rm}{\normalfont\rmfamily}{\mathrm}
\DeclareOldFontCommand{\sf}{\normalfont\sffamily}{\mathsf}
\DeclareOldFontCommand{\tt}{\normalfont\ttfamily}{\mathtt}
\DeclareOldFontCommand{\bf}{\normalfont\bfseries}{\mathbf}
\DeclareOldFontCommand{\it}{\normalfont\itshape}{\mathit}
\DeclareOldFontCommand{\sl}{\normalfont\slshape}{\@nomath\sl}
\DeclareOldFontCommand{\sc}{\normalfont\scshape}{\@nomath\sc}
\makeatother
\usepackage{amsmath,amssymb,graphicx,scalefnt,enumitem}
\usepackage[absolute]{textpos}
\usepackage[noadjust]{cite}
\usepackage[affil-it]{authblk}
\usepackage{xcolor}
\usepackage{tabularx,booktabs}
\usepackage{xspace}
\usepackage{listings}
\usepackage{algorithm}
\usepackage{pifont}
\usepackage{algorithmicx}
\usepackage{algpseudocode}
\usepackage[final]{showkeys}
\usepackage{filemod}
\usepackage{bold-extra}
\usepackage[font=small,labelfont=bf,format=plain,margin=0.05\textwidth]{caption}
\usepackage{subfig}

\newcommand{\cmark}{\ding{51}}%
\newcommand{\xmark}{\ding{55}}%
\usepackage[pdftitle={Reconstructing Rational Functions with FireFly},
  pdfauthor={Jonas Klappert and Fabian Lange},
  pdfkeywords={Finite Field, functional reconstruction, functional interpolation},
  bookmarks=true, linktocpage,
  colorlinks=true, allbordercolors=white, allcolors=blue]{hyperref}
\newcommand{\pheaderline}{{\footnotesize TTK-19-11\\P3H-19-007}}
\newcounter{notecount}

\newcommand{\citere}[1]{Ref.\,\cite{#1}}
\newcommand{\citeres}[1]{Refs.\,\cite{#1}}
\newcommand{\code}[1]{\texttt{#1}}
\newcommand{\abbrev}[1]{{\scalefont{1}#1}}
\newcommand{\alg}[1]{Alg.\,(\ref{#1})}
\newcommand{\eqn}[1]{Eq.\,(\ref{#1})}

\newcommand{\fig}[1]{Fig.\,\ref{#1}}

\newcommand{\tab}[1]{Tab.\,\ref{#1}}

\newcommand{\sct}[1]{Sect.\,\ref{#1}}
\newcommand{\app}[1]{App.\,\ref{#1}}

\newcommand{\dd}{{\rm d}}

\newcommand{\vecz}{$\vec{z} = (z_1,\dots z_n)$}
\newcommand{\zp}[1]{{\mathbb{Z}}_{#1}}

\def\cpc#1#2#3#4{\href{https://dx.doi.org/#4}{{\it Comp.~Phys.~Commun.~}\jref{\bf #1}{#2}{#3}}}

\def\ijmpa#1#2#3#4{\href{https://dx.doi.org/#4}{{\it Int.~J.~Mod.~Phys.~}\jref{\bf A #1}{#2}{#3}}}

\def\jacm#1#2#3#4{\href{https://dx.doi.org/#4}{{\it J.~ACM~}\jref{\bf #1}{#2}{#3}}}

\def\jhep#1#2#3#4{\href{https://dx.doi.org/#4}{{\it JHEP~}\jref{\bf #1}{#2}{#3}}}
\def\jpcs#1#2#3#4{\href{https://dx.doi.org/#4}{{\it J.~Phys.~Conf.~Ser.~}\jref{\bf #1}{#2}{#3}}}

\def\npb#1#2#3#4{\href{https://dx.doi.org/#4}{{\it Nucl.~Phys.~}\jref{\bf B #1}{#2}{#3}}}
\def\npps#1#2#3#4{\href{https://dx.doi.org/#4}{{\it Nucl.~Phys.~Proc.~Suppl.~}\jref{\bf #1}{#2}{#3}}}
\def\plb#1#2#3#4{\href{https://dx.doi.org/#4}{{\it Phys.~Lett.~}\jref{\bf B #1}{#2}{#3}}}

\def\prd#1#2#3#4{\href{https://dx.doi.org/#4}{{\it Phys.~Rev.~}\jref{\bf D #1}{#2}{#3}}}

\def\prl#1#2#3#4{\href{https://dx.doi.org/#4}{{\it Phys.~Rev.~Lett.~}\jref{\bf #1}{#2}{#3}}}

\def\zpc#1#2#3#4{\href{https://dx.doi.org/#4}{{\it Z.~Phys.~}\jref{\bf C #1}{#2}{#3}}}

\def\esam#1#2#3#4{\href{https://dx.doi.org/#4}{{\it Symbolic Algebraic Comp.\ EUROSAM~}\jref{\bf #1}{#2}{#3}}}
\def\focs#1#2#3#4{\href{https://dx.doi.org/#4}{{\it Proc.\ Symp.\ Foundations Comp.\ Sci.~}\jref{\bf #1}{#2}{#3}}}
\def\issacfirst#1#2#3#4{\href{https://dx.doi.org/#4}{{\it Symbolic Algebraic Comp.\ ISSAC~}\jref{\bf #1}{#2}{#3}}}
\def\issac#1#2#3#4{\href{https://dx.doi.org/#4}{{\it Proc.\ Int.\ Symp.\ Symbolic Algebraic~Comp.~}\jref{\bf #1}{#2}{#3}}}
\def\icms#1#2#3#4{\href{https://dx.doi.org/#4}{{\it Proc.\ Int.\ Congress Mathe.\ Software~}\jref{\bf #1}{#2}{#3}}}
\def\jsc#1#2#3#4{\href{https://dx.doi.org/#4}{{\it J.~Symb.~Comp.~}\jref{\bf #1}{#2}{#3}}}
\def\macis#1#2#3#4{\href{https://dx.doi.org/#4}{{\it Math.\ Aspects Comp.\ Information Sci.~}\jref{\bf #1}{#2}{#3}}}

\def\pasco#1#2#3#4{\href{https://dx.doi.org/#4}{{\it Proc.\ Int.\ Workshop Parallel Symbolic Comp.~}\jref{\bf #1}{#2}{#3}}}
\def\sigsamb#1#2#3#4{\href{https://dx.doi.org/#4}{{\it ACM SIGSAM Bulletin~}\jref{\bf #1}{#2}{#3}}}
\def\smjcat#1#2#3#4{\href{https://dx.doi.org/#4}{{\it SIAM~J.~Comp.~}\jref{\bf #1}{#2}{#3}}}
\def\stoc#1#2#3#4{\href{https://dx.doi.org/#4}{{\it Proc.\ ACM Symp.\ Theory Comp.~}\jref{\bf #1}{#2}{#3}}}
\def\symsac#1#2#3#4{\href{https://dx.doi.org/#4}{{\it Proc.\ ACM Symp.\ Symbolic Algebraic Comp.~}\jref{\bf #1}{#2}{#3}}}
\def\tcs#1#2#3#4{\href{https://dx.doi.org/#4}{{\it Theor.~Comp.\ Sci.~}\jref{\bf #1}{#2}{#3}}}
\def\otherjournal#1#2#3#4#5{\href{https://dx.doi.org/#5}{{\it #1}\jref{\bf #2}{#3}{#4}}}
\newcommand{\jref}[3]{{\bf #1} (#2) #3}
\newcommand{\hepph}[1]{\href{https://arXiv.org/abs/hep-ph/#1}{\texttt{hep-ph/#1}}}

\newcommand{\arxiv}[2]{\href{https://arXiv.org/abs/#1}{\texttt{arXiv:#1\,[#2]}}}
\newcommand{\bibentry}[4]{#1, {\it #2}, #3\ifthenelse{\equal{#4}{}}{}{, }#4.}

\title{Reconstructing Rational Functions with \code{FireFly}}
\author{Jonas Klappert and Fabian Lange}
\affil{Institute for Theoretical Particle Physics and Cosmology, RWTH
  Aachen University, D-52056 Aachen, Germany}
\date{}
\usepackage[parfill]{parskip}
\begin{document}
\maketitle
\thispagestyle{empty}
\begin{abstract}
We present the open-source \code{C++} library \code{FireFly} for the reconstruction of multivariate rational functions over finite fields.
We discuss the involved algorithms and their implementation.
As an application, we use \code{FireFly} in the context of integration-by-parts reductions and compare runtime and memory consumption to a fully algebraic approach with the program \code{Kira}.
\end{abstract}

\begin{textblock*}{10em}(\textwidth,1.5cm)
\raggedright\noindent
\pheaderline
\end{textblock*}

\clearpage
\section*{PROGRAM SUMMARY}
\textit{Manuscript Title:} Reconstructing Rational Functions with \code{FireFly}\\
\textit{Authors:} Jonas Klappert and Fabian Lange\\
\textit{Program title:} \code{FireFly}\\
\textit{Licensing provisions:} \abbrev{GNU} General Public License 3 (\abbrev{GPL})\\
\textit{Programming language:} \code{C++}\\
\textit{Computer(s) for which the program has been designed:} Desktop PC, compute nodes/workstations\\
\textit{Operating system(s) for which the program has been designed:} Linux 64-bit\\
\textit{RAM required to execute with typical data:} Depends on the complexity of the problem, ranging from few MiB to few hundred GiB.\\
\textit{Has the code been vectorized or parallelized?:} Yes\\
\textit{Number of processors used:} Any number of cores\\
\textit{Supplementary material:} This article\\
\textit{Keywords:} finite field, black-box interpolation, modular arithmetic, symbolic calculations\\
\textit{CPC Library Classification:} 4.8 Linear Equations and Matrices, 4.10 Interpolation, 4.12 Other Numerical Methods\\
\textit{External routines/libraries used:} \abbrev{GNU GMP}\,[1], \abbrev{FLINT}\,[2]\\
\textit{Nature of problem:} The interpolation of an unknown rational function, called black box, from only its evaluations can be used in many physical contexts where algebraic calculations fail due to memory and time restrictions.\\
\textit{Solution method:} The black-box function is evaluated at different points over a finite field. These points are then used by interpolation algorithms\,[3,4] to obtain the analytic form of the black-box function. The members of a finite field are promoted to $\mathbb{Q}$ using a rational reconstruction algorithm\,[5,6].\\
\textit{Restrictions:} The \abbrev{CPU} time and the available \abbrev{RAM}\\
\textit{Running time:} Milliseconds to weeks, depending on the complexity of the problem\\
\textit{References:}\\
$[1]$ T.~Granlund et~al., \textit{The GNU Multiple Precision Arithmetic Library},\\\href{https://gmplib.org/}{\texttt{https://gmplib.org/}}.\\
$[2]$ W.~Hart et al., \textit{FLINT: Fast Library for Number Theory}, \href{http://www.flintlib.org/}{\texttt{http://www.flintlib.org/}}.\\
$[3]$ R.~Zippel, \textit{Interpolating Polynomials from their Values}, \jsc{9}{1990}{375--403}{10.1016/S0747-7171(08)80018-1}.\\
$[4]$ A.~Cuyt, W.-s.~Lee, \textit{Sparse interpolation of multivariate rational functions}, \tcs{412}{2011}{1445--1456}{10.1016/j.tcs.2010.11.050}.\\
$[5]$ P.S.~Wang, \textit{A p-adic Algorithm for Univariate Partial Fractions}, \symsac{1981}{1981}{212--217}{10.1145/800206.806398}.\\
$[6]$ M.~Monagan, \textit{Maximal Quotient Rational Reconstruction: An Almost Optimal Algorithm for Rational Reconstruction}, \issac{2004}{2004}{243--249}{10.1145/1005285.1005321}.
\clearpage
\tableofcontents
\clearpage

\section{Introduction}
The interpolation of polynomials has been researched since the 18th century and some of the algorithms, e.g.\ Newton's interpolation algorithm\,\cite{Abramowitz}, are still used today.
They use black-box probes, i.e.\ evaluations at numerical values, of the to be interpolated function, the black box, for the interpolation.
However, these original algorithms were designed for univariate polynomials and are costly to generalize to the multivariate case.
Interpolation algorithms specific for multivariate polynomials have been studied intensively for several decades, e.g.\ in \citeres{Zippel:1979,Zippel:1990,Ben-Or:1988,Kaltofen:1988,Kaltofen:1990_2,Kaltofen:2000,Kaltofen:2003,Javadi:2010}.

The interpolation of rational functions is a younger field of research, even though Thiele's univariate interpolation formula\,\cite{Abramowitz} has been known for more than one hundred years.
To our knowledge, the first multivariate algorithms came up in the early 1990s.
In 1990, Kaltofen and Trager presented an algorithm to evaluate numerator and denominator of a rational function separately and suggested to combine it with a sparse polynomial interpolation to interpolate a multivariate rational function\,\cite{Kaltofen:1990_1}.
Kaltofen and Yang refined this algorithm in 2007\,\cite{Kaltofen:2007}.
Also in 1990, Grigoriev, Karpinski, and Singer presented an algorithm to obtain a bound on the degrees of numerator and denominator, then build a system of equations, and solve it for all possible coefficients\,\cite{Grigoriev:1990,Grigoriev:1991,Grigoriev:1994}.
This is a dense algorithm, i.e.\ it assumes that all coefficients of the rational function up to the bounds are non-zero.
Another algorithm was published by de Kleine, Monagan, and Wittkopf in 2005\,\cite{deKleine:2005}.
It is based on a modified version of the Extended Euclidean Algorithm\,\cite{von_zur_Gathen,Monagan:2004}.
The algorithm by Cuyt and Lee from 2011 first performs a univariate interpolation of the rational function and then uses the coefficients as input for a multivariate polynomial interpolation\,\cite{Cuyt:2011}.
In 2017, Huang and Gao\,\cite{Huang:2017} presented an algorithm which uses Kronecker substitution\,\cite{von_zur_Gathen}.

Most of the algorithms mentioned above rely on finite field arithmetic.
Finite fields, e.g.\ integers modulo a prime number, are used for several hundred years.
These calculations are done in an exact arithmetic environment which, however, avoids a number swell.
This means that the coefficients of the polynomials and rational functions obtained by the interpolation algorithms are numbers in the finite field.
In 1981, Wang presented an algorithm based on the Extended Euclidean Algorithm\,\cite{von_zur_Gathen} which allows to reconstruct a rational number from its image modulo an integer\,\cite{Wang:1981}.

We review finite fields, interpolation algorithms for both polynomials and rational functions, and the reconstruction algorithms for rational numbers in \sct{sec:functional_interpolation}.
The \code{C++} library \code{FireFly} uses some of these algorithms.
For the rational function reconstruction we employ the algorithm by Cuyt and Lee\,\cite{Cuyt:2011} combined with Zippel's algorithm for the polynomial interpolation\,\cite{Zippel:1979,Zippel:1990}.
We also present some modifications to reduce the number of black-box probes.
In \sct{sec:firefly} we present the implementation of \code{FireFly} and show how it can be used.
We also present some benchmarks for the interpolation of multivariate rational functions.

As an application we apply \code{FireFly} to the reduction of Feynman integrals with integration-by-parts (\abbrev{IBP}) relations\,\cite{Tkachov:1981wb,Chetyrkin:1981qh}, which are a standard calculation technique in multi-loop calculations of theoretical particle physics.
They relate different integrals with each other through linear relations.
The first solution strategy is to rearrange the \abbrev{IBP} relations to recursion relations, which always express an integral through easier integrals.
Applying the recursion relations repeatedly allows one to express all integrals through master integrals.
At three-loop level, this approach has been successfully applied for massless propagator-type diagrams\,\cite{Chetyrkin:1981qh,Gorishnii:1989gt,Larin:1991fz}, massive tadpoles\,\cite{Broadhurst:1991fi,Steinhauser:2000ry}, and on-shell propagators\,\cite{Melnikov:1999xp,Melnikov:2000qh,Melnikov:2000zc}.
However, at higher orders in perturbation theory it becomes impractical to obtain such recursion relations.
Nonetheless, some progress in deriving such relations automatically has been made in the last decade\,\cite{Lee:2012cn,Smirnov:2013dia,Lee:2013mka,Ruijl:2017cx}.

The second strategy was presented by Laporta in 2001\,\cite{Laporta:2001dd}.
He suggested one builds a system of equations out of the \abbrev{IBP} relations by inserting values for the propagator powers and then solves this system.
The Laporta algorithm and its modifications have been implemented in several public codes, \code{AIR}\,\cite{Anastasiou:2004vj}, \code{FIRE}\,\cite{Smirnov:2008iw,Smirnov:2013dia,Smirnov:2014hma,Smirnov:2019qkx}, \code{Reduze}\,\cite{Studerus:2009ye,vonManteuffel:2012np}, and \code{Kira}\,\cite{Maierhoefer:2017hyi,Maierhofer:2018gpa}, as well as in numerous private codes.
However, the systems of equations can become gigantic and, thus, expensive to solve both in terms of memory and runtime, which is partly related to large intermediate expressions.

To circumvent these problems, the use of finite-field techniques has been proposed by Kauers in 2008 together with a \code{Mathematica} package as an example\,\cite{Kauers:2008zz}.
In 2013, Kant proposed to solve the system of equations over a finite field before running the Laporta algorithm in order to reduce the size of the system by identifying and removing the linearly dependent equations\,\cite{Kant:2013vta}.
This auxiliary use of finite fields has been implemented in \code{Kira} and led to significantly improved runtimes\,\cite{Maierhoefer:2017hyi}.
The use of the interpolation techniques over finite fields in the context of the Laporta algorithm has been advocated in \citere{vonManteuffel:2014ixa}.
In 2016, Peraro summarized some of the interpolation techniques from computer science in the context of generalized unitarity\,\cite{Peraro:2016wsq}.
The first calculation using finite-field interpolation techniques for \abbrev{IBP} reductions was accomplished by von Manteuffel and Schabinger in 2016\,\cite{vonManteuffel:2016xki}.
This was a one scale problem and, thus, also a one variable problem.
Recently, three more one-scale calculations have been finished\,\cite{Henn:2019rmi,vonManteuffel:2019wbj,vonManteuffel:2019gpr}.
The recently published version 1.2 of \code{Kira} uses a multivariate Newton interpolation as supportive technique\,\cite{Maierhofer:2018gpa}, albeit over $\mathbb{Z}$ instead of a finite field.
Shortly after, \code{FIRE6} was published as first public implementation of a Laporta algorithm with a multivariate interpolation of rational functions over a finite field\,\cite{Smirnov:2019qkx}.
It supports the stable interpolation with two variables, i.e.\ two scales when setting one of them to one.

After a concise review of \abbrev{IBP} reductions, we briefly describe our implementation with \code{FireFly} and compare it with the traditional reduction with \code{Kira} for some examples in \sct{sec:IBP}.
We also point out the advantages and disadvantages of the finite-fields-interpolation approach.

Another important method for multi-loop calculations is based on generalized unitarity.
As mentioned above, Peraro pioneered the application of finite-field interpolation techniques for this method in 2016\,\cite{Peraro:2016wsq}.
This lead to several successful calculations in the last two years\,\cite{Badger:2017jhb,Abreu:2017hqn,Bern:2018jmv,Abreu:2018jgq,Badger:2018enw,Abreu:2018zmy,Abreu:2018aqd,Abreu:2019rpt}, where functions with up to four variables have been interpolated.
However, we do not delve into this field in our paper.

\section{Functional interpolation over finite fields}
\label{sec:functional_interpolation}
The interpolation of a function $f$ of $n$ variables \vecz{} is based on the idea to require no knowledge about the function itself and only use its evaluations at different values of $\vec{z}$ to construct an analytic form of the probed function or at least find an approximation formula.
These kinds of interpolation problems are also called black-box interpolation problems, where the function to be interpolated serves as the black box.
These interpolations are performed over a field. A set of particularly suited fields, due to their specific properties, are finite fields $\zp{p}$ with characteristic $p$, where $p$ is the defining prime, on which we will focus in this paper. All calculations are thus carried out modulo $p$ avoiding number swell and thus saving memory and runtime. The multiplicative inverse in $\zp{p}$ is unique and can be determined using the Extended Euclidean Algorithm\,\cite{von_zur_Gathen}.

In the following sections, we describe how polynomials and rational functions
of in principle arbitrarily many variables can be efficiently interpolated and how one can
promote elements of $\zp{p}$ to the field of rational numbers $\mathbb{Q}$.

\subsection{Interpolation of polynomials}
\label{ssec:int_poly}
To fix the notation, we start by defining multivariate polynomials as follows.
Given a set of $n$ variables \vecz{} and an $n$-dimensional multi-index
$\alpha=(\alpha[1],\dots,\alpha[n])$ containing integers $\alpha[i] \ge 0$,
we define a monomial $\vec{z}^{\;\alpha}$ as
\begin{equation}
\vec{z}^{\;\alpha} \equiv \prod_{i=1}^n z_i^{\alpha[i]}
\end{equation}
with a degree $d$ of
\begin{equation}
d = \sum_{i=1}^n \alpha[i].
\end{equation}
A polynomial $f$, which is a member of the polynomial ring $\zp{p}[\vec{z}]$
in the variables $\vec{z}$, is defined as
\begin{equation}
f(\vec{z}) = \sum_{j=1}^{T} c_{\alpha_j} \vec{z}^{\;\alpha_j} ,
\label{eq:poly_canonical}
\end{equation}
where $T$ is the number of non-zero terms.
The coefficients $c_{\alpha_j}$ are elements of $\zp{p}$ corresponding to different multi-indices $\alpha_j$.

Solving the black-box interpolation problem of a multivariate polynomial can be done by recursive interpolations
of univariate polynomials. Thus, we briefly mention how to interpolate
univariate polynomials before turning to the multivariate case. A well-known
interpolation algorithm for univariate polynomial functions $f=f(z_1)\in \zp{p}[z_1]$ is given
by the Newton interpolation which relies on Newton polynomials\,\cite{Abramowitz}.
Given a sequence of distinct interpolation points $y_{1,1},\dots,y_{1,D+1} \in \zp{p}$, the Newton
polynomial for $f$ of degree $D$ can be written as
\begin{equation}
\label{eq:newton}
f(z_1) = a_0 + \sum_{i=1}^{D} a_i \prod_{j=1}^{i}(z_1 - y_{1,j}),
\end{equation}
where the coefficients $a_i$ can be recursively defined as
\begin{align}
a_i & \equiv a_{i,i},\\
a_{i,j} & = \frac{a_{i,j-1} - a_{j-1}}{y_{1,i + 1} - y_{1,j}},\\
a_{i,0} & = f(y_{1,i + 1}).
\end{align}
A benefit of this algorithm is that the calculation of a new term does not alter the previously computed terms.
This feature makes the Newton interpolation particularly suited for interpolations
with an unknown degree $D$.

In principle, Newton's algorithm computes infinitely many $a_i$ and one can stop the interpolation if the desired accuracy is reached.
However, we utilize an early termination approach which is based on the assumption that if a given number of $a_i$ vanish we have correctly interpolated the black-box function with high probability.
This probability can be quantified in the following way.
Given distinct evaluation points for a black box $f$, a field $\zp{p}$, a positive integer threshold $\eta$, and the smallest non-negative integer $D$ such that
\begin{equation}
a_D = a_{D+1} = \ldots = a_{D+\eta} = 0,
\end{equation}
then the probability of this approach to succeed is no less than \cite{Kaltofen:2000,Kaltofen:2003}
\begin{equation}
1 - (D+1)\left(\frac{D}{p}\right)^\eta.
\end{equation}
The probability increases with the number of additional $a_i$ computed for random choices of $y_{1,i}$.
It can also be increased by choosing a larger characteristic for $\zp{p}$.
In our studies, we found no wrongly terminated interpolation with $\eta=1$ and defining $\zp{p}$ with the largest 63-bit primes.
Hence, for practical usage, it might be sufficient to set $\eta=1$.

The algorithm can be easily generalized to the case of multivariate polynomials by a recursive
application of the Newton interpolation for univariate functions, e.g.\ as proposed in \citere{Peraro:2016wsq}. Consider a generic
multivariate polynomial $f\in \zp{p}[z_1,\dots,z_n]$. $f$ can be reinterpreted as a univariate polynomial in $z_1$ of degree $D$, with
its coefficients being multivariate polynomials of $n-1$ variables $(z_2,\dots,z_n)$. This
promotes the coefficients $a_i$ to be elements of the polynomial ring $\zp{p}[z_2,\dots z_n]$
such that
\begin{equation}
f(z_1,\dots,z_n) = \sum_{i=0}^{D} a_i(z_2,\dots,z_{n})\prod_{j=1}^{i}(z_1 - y_{1,j}),
\end{equation}
where $y_{1,i} \in \zp{p}$ are randomly chosen interpolation points for $z_1$ as in the univariate case.
Obviously, one can again use the same partition for $a_i(z_2,\dots,z_{n})$ leading to new
coefficients which are functions of $n-2$ arguments. After applying this procedure $n-1$
times, we are left with the univariate interpolation problem, which can be solved
by the Newton interpolation. For later usage, we denote the numerical choices $y_{i,j}$ for $z_i$ as the corresponding value to $z_i$ at order $j$. The orders of each variable form a tuple which we call $z_i$ order, i.e.\ the numerical choices
$y_{2,2}$, $y_{3,3}$ form the $z_i$ order $(2,3)$.

For practicality, it is useful to transform the interpolated Newton polynomial
to its canonical form defined in \eqn{eq:poly_canonical}. This can be done by using additions
of multivariate polynomials and multiplications of a multivariate polynomial by a linear
univariate polynomial.

However, the Newton interpolation algorithm is a \textit{dense} algorithm, i.e.\ it interpolates all possible terms of a polynomial even if it is \textit{sparse} and only a small subset of coefficients is non-zero.
It requires $\prod_i (D_i+2)$ black-box probes of the polynomial $f(z_1,\dots,z_n)$ in general, where $D_i$ is the maximal degree of $z_i$. Overall, the complexity of this algorithm is exponential in $n$ and grows quickly as the number of variables and the individual maximal degrees increase, which is inefficient for sparse polynomials. Note that a degree bound on each variable can reduce the complexity to be $\prod_i (D_i+1)$. This is still not preferable for sparse polynomials which are usually encountered in physical calculations.

A more efficient multivariate algorithm in terms of black-box probes is the Zippel algorithm\,\cite{Zippel:1979,Zippel:1990}, which in addition takes advantage of the sparsity of the polynomial.
As in the previous algorithm, Zippel's algorithm interpolates one variable at a time.
The interpolation of a variable is called a \textit{stage}, i.e.\ stage one is the interpolation $f(z_1, y_{2,1}, y_{3,1}, \dots)$, stage two the interpolation $f(z_1, z_2, y_{3,1}, \dots)$ and so on.
After the first stage, one interpolates each coefficient of the previous stage as a univariate polynomial.
The main advantage of the Zippel algorithm is the following.
If a polynomial coefficient $c_\alpha$ evaluates to zero at one stage, one assumes that it will also vanish at all other stages.
This can avoid a sizable amount of black-box probes compared to a dense interpolator.
Since this assumption is probabilistic, it is crucial to choose the numbers at which the polynomial is probed with great care in order for this assumption to hold. To make this choice more robust, one usually uses seed numbers which are called \textit{anchor points}. The anchor points should be set to random numbers to minimize the chance of coincidental cancellations. All other $z_i$ order values can be obtained by computing the corresponding power of the anchor point, i.e.\ if $y_{i,1}=2$ then the corresponding $j$th $z_i$ order value is $y_{i,j} = 2^j$.

For illustration we provide an example of the Zippel algorithm.
Consider the following polynomial to be interpolated
\begin{equation}
  f(z_1, z_2, z_3) = c_{\alpha_1} z_1^5 + c_{\alpha_2} z_1z_2^4 + c_{\alpha_3} z_1z_2z_3^3 + c_{\alpha_4} z_2^5.
  \label{eq:Zippel_ex}
\end{equation}
The index $\alpha_i$ is the multi-index of the corresponding monomial, i.e.\ $\alpha_1 = (5,0,0)$.
In the first stage, one interpolates the univariate polynomial in $z_1$ with Newton's algorithm by fixing $z_2 = y_{2,1}$ and $z_3 = y_{3,1}$.
This yields
\begin{equation}
  f(z_1, y_{2,1}, y_{3,1}) = k_0(y_{2,1}, y_{3,1}) + k_1(y_{2,1}, y_{3,1}) \cdot z_1 + k_5(y_{2,1}, y_{3,1}) \cdot z_1^5
  \label{eq:stage1}
\end{equation}
after six black-box probes and rewriting the Newton polynomial in canonical form. An additional probe is needed to verify the termination of this stage.
The polynomial coefficients $k_i(y_{2,1}, y_{3,1})$, not to be confused with the coefficients $a_i$ of Newton's interpolation, are multivariate polynomials in the remaining variables $z_2$ and $z_3$ evaluated at $y_{2,1}$ and $y_{3,1}$.

At stage two, the coefficients $k_i(y_{2,1}, y_{3,1})$ are promoted to polynomials $k_i(z_2,y_{3,1})$.
Since $k_2 = k_3 = k_4 = 0$, we also assume the corresponding polynomials $k_i(z_2,z_3)$ to vanish.
Therefore, only three polynomials have to be interpolated in $z_2$ at stage two.
The univariate interpolation in $z_2$ is done by using the numerical values of $k_i(y_{2,j}, y_{3,1})$ of the polynomials
\begin{equation}
  f_j(z_1, y_{2,j}, y_{3,1}) = k_{0}(y_{2,j}, y_{3,1}) + k_{1}(y_{2,j}, y_{3,1}) \cdot z_1 + k_{5}(y_{2,j}, y_{3,1}) \cdot z_1^5
\end{equation}
as black-box probe for the univariate interpolation of $k_i(z_2,y_{3,1})$.
The result of the first stage, $f_1(z_1, y_{2,1}, y_{3,1})$ given by \eqn{eq:stage1}, can be reused.
The other polynomials $f_j(z_1, y_{2,j}, y_{3,1})$ can again be interpolated by a univariate Newton interpolation.
However, since we already know that only three coefficients contribute to the univariate polynomial $f_j(z_1, y_{2,j}, y_{3,1})$, it is more efficient to build a system of three equations and solve it.
Zippel's original algorithm requests new polynomials $f_j(z_1, y_{2,j}, y_{3,1})$ until all $k_i(z_2,y_{3,1})$ are interpolated, i.e.\ six times in the example.
This means that stage two would require $6 \times 3 = 18$ black-box probes in total.

Kaltofen, Lee, and Lobo suggested to remove those $k_i(z_2,y_{3,1})$ from the system where the Newton interpolation terminates, which reduces the size of the system for the remaining coefficients\,\cite{Kaltofen:2000,Kaltofen:2003}.
This procedure is called \textit{temporary pruning}.
In the example, the interpolation of $k_5(z_2,y_{3,1})$ terminates after the second step, the evaluation of $k_1(z_2,y_{3,1})$ after the sixth, and $k_0(z_2,y_{3,1})$ after the seventh, which corresponds to $3 + 4 \times 2 + 1 = 12$ black-box probes for stage two.

The result of stage two is
\begin{equation}
  \begin{split}
    f(z_1, z_2, y_{3,1}) &= {\tilde k_0}(y_{3,1}) \cdot z_2^5 + ({\tilde k_1}(y_{3,1}) \cdot z_2 + {\tilde k_2}(y_{3,1}) \cdot z_2^4) z_1 + {\tilde k_5}(y_{3,1}) \cdot z_1^5 \\
    &= {\tilde k_0}(y_{3,1}) \cdot z_2^5 + {\tilde k_1}(y_{3,1}) \cdot z_1 z_2 + {\tilde k_2}(y_{3,1}) \cdot z_1 z_2^4 + {\tilde k_5}(y_{3,1}) \cdot z_1^5 .
  \end{split}
  \label{eq:stage2}
\end{equation}
Stage three starts by promoting ${\tilde k_i}(y_{3,1})$ to polynomials ${\tilde k_i}(z_3)$ and proceeds exactly as stage two.
The interpolations of ${\tilde k_0}(z_3)$, ${\tilde k_2}(z_3)$, and ${\tilde k_5}(z_3)$ terminate after the second step, and the interpolation of ${\tilde k_1}(z_3)$ after the fifth.
Thus, Zippel's original algorithm requires $4 \times 4 = 16$ and the improved version $4 + 3 \times 1 = 7$ black-box probes for stage three.

Therefore, the complete interpolation of \eqn{eq:Zippel_ex} requires $7 + 18 + 16 = 41$ black-box probes with the original version and $7 + 12 + 7 = 26$ with the improved version using temporary pruning.
In contrast, the recursive multivariate Newton interpolation needs $(5 + 2) \times (5 + 2) \times (3 + 2) = 245$ probes.

The Zippel algorithm can be further improved by assuming knowledge of the maximal degree $D$, which always is the case when using it as part of the multivariate interpolation of rational functions as described in \sct{sec:int_rat}.
Then, one can abort the univariate Newton interpolation of $k_i$ after step $j$ if $j = D - d_i$, where $d_i$ is the total degree of the monomial corresponding to $k_i$.
This procedure is called \textit{permanent pruning}\,\cite{Diaz:1998} and it reduces the total number of black-box probes to 20 for the given example, whereas the multivariate Newton algorithm only drops to $(5 + 1) \times (5 + 1) \times (3 + 2) = 180$ probes.
In contrast, there are $\binom{3 + 5}{5} = 56$ possible coefficients and solving for them with a system of equations requires the same amount of black-box probes.
This clearly shows the advantage of the Zippel algorithm compared to a recursive multivariate Newton interpolation.

Furthermore, if we know that all monomials have the same degree $d_i = 5$, the example already finishes after stage two.
From \eqn{eq:stage2} we can deduce that $c_{\alpha_3} = \tilde k_1(y_{3,1}) / y_{3,1}^3$ is the coefficient for $z_1z_2z_3^3$.
This prediction of the remaining powers is not used in our implementation of the Zippel algorithm, because it is closely related to the homogenization procedure\,\cite{Diaz:1998} which will be done for the rational function interpolation in \sct{sec:int_rat}.

An additional optimization regarding runtime can be achieved by noticing that the systems of equations which have to be solved during each stage in Zippel's algorithm are generalized transposed Vandermonde systems\,\cite{Zippel:1990}
\begin{equation}
\begin{pmatrix}
  1&1&\dots&1\\
  v_{\alpha_1}&v_{\alpha_2}&\dots&v_{\alpha_T}\\
\vdots&\vdots&\ddots&\vdots\\
v_{\alpha_1}^{T-1} & v_{\alpha_2}^{T-1}&\dots&v_{\alpha_T}^{T-1}
\end{pmatrix}
\begin{pmatrix}
c_{\alpha_1}\\
c_{\alpha_2}\\
\vdots\\
c_{\alpha_T}
\end{pmatrix}
=\begin{pmatrix}
f(\vec{y}^{\;0})\\
f(\vec{y}^{\;1})\\
\vdots\\
f(\vec{y}^{\;T-1})
\end{pmatrix},
\label{eq:vander_1}
\end{equation}
where $f(\vec{z})$ is a multivariate polynomial of n variables with $T$ terms,
\begin{equation}
 v_{\alpha_i} = \vec{y}^{\;\alpha_i} = \prod_{j=1}^n y_{j,1}^{\alpha_i[j]},\qquad \vec{y}^{\;i}= \{y_{1,1}^i,y_{2,1}^i,\ldots,y_{n,1}^i\}.
 \end{equation}
Note that a Vandermonde system assumes that the evaluation points of $f$ are powers of the anchor points $\vec{y} = \{y_{1,1},y_{2,1},\ldots,y_{n,1}\}$ and are not chosen randomly apart from the anchor points themselves.
All $v_{\alpha_i}$ have to be increasing monotonically with respect to $\alpha_i$, with ${\alpha_1}$ being the lowest degree.
In order to have a unique solution of \eqn{eq:vander_1}, we require that the monomial evaluations $v_{\alpha_i}$ are distinct in $\zp{p}$, i.e. for a $T\times T$ system we need $T$ distinct values from $\zp{p}$.
Solving algorithms for Vandermonde systems are much more efficient compared to Gaussian elimination, since Vandermonde systems can be solved using only $O(T^2)$ time and $O(T)$ space\,\cite{Zippel:1990,Kaltofen:1988}. For usual Vandermonde systems presented in \eqn{eq:vander_1}, the first evaluation of $f(\vec{z})$ will be done while setting all variables equal to 1.
This variable choice increases the probability that monomials cancel, which is especially dangerous if $f(\vec{z})$ is the denominator of a rational function.
To circumvent this problem, we consider a shifted version of generalized transposed Vandermonde systems
\begin{equation}
\begin{pmatrix}
  v_{\alpha_1}&v_{\alpha_2}&\dots&v_{\alpha_T}\\
  v_{\alpha_1}^2&v_{\alpha_2}^2&\dots&v_{\alpha_T}^2\\
\vdots&\vdots&\ddots&\vdots\\
v_{\alpha_1}^{T} & v_{\alpha_2}^{T}&\dots&v_{\alpha_T}^{T}
\end{pmatrix}
\begin{pmatrix}
c_{\alpha_1}\\
c_{\alpha_2}\\
\vdots\\
c_{\alpha_T}
\end{pmatrix}
=\begin{pmatrix}
f(\vec{y}^{\;1})\\
f(\vec{y}^{\;2})\\
\vdots\\
f(\vec{y}^{\;T})
\end{pmatrix} ,
\label{eq:vandermonde}
\end{equation}
where all exponents in the matrix are increased by one.
Systems with a general shift of the exponents were considered in \citere{Hu:2016}.
These systems preserve all properties of general Vandermonde systems and only require a modification of the solving algorithm, while making cancellations very unlikely. An algorithm to solve such systems is given in \app{sec:algorithms}.

Note that it is beneficial in terms of the number of black-box probes to choose the variable order descending in the maximal degrees, i.e.\ that the lowest degree variable should be the last one to interpolate and the highest the first.
This follows from the nature of the Zippel algorithm in which intermediate systems of equations usually grow with every stage.
Generally, the best variable order for a black-box function is not known a priori, but for physical calculations one can utilize problems which are similar but simpler to obtain a guess on how the individual degrees could be distributed. We will illustrate the impact on the number of black-box probes of such a choice in \sct{ssec:Benchmarks}.

Assuming a total degree bound $D$, the Zippel algorithm scales as $O(nDT)$, with $T$ being the number of non-zero coefficients\,\cite{Zippel:1990}.
In the completely dense case, the improved version with pruning requires $\binom{n+D}{D}$ probes with a degree bound, which is exactly one probe for every coefficient.
Without a degree bound, only a small number of additional probes is required to verify that the interpolation terminates with high probability.
However, there is a small chance that the interpolated polynomial of this algorithm is wrong. This can happen if a bad combination of anchor points is chosen.
Zippel proved that if $y_{i,1}$ are chosen uniformly randomly from a field $\zp{p}$, the probability that the interpolation of a black box $f$ with $n$ variables, degree $D$, and non-zero terms $T$ fails is less than \cite{Zippel:1990}
\begin{equation}
\frac{nD^2T^2}{p} .
\end{equation}
For this bound one assumes that no singular Vandermonde systems arise during the interpolation and all zeros are avoided by the choice of anchor points. It is based on the Zippel-Schwartz lemma \cite{Zippel:1979,Schwartz:1980}
\begin{equation}
\mathrm{Pr}[f(\vec{y}) = 0] \le \frac{D}{|S|} ,
\end{equation}
which provides a bound on the probability (Pr) that a polynomial $f$ of total degree $D$ evaluates to zero when selecting $\vec{y}$ independently and uniformly randomly from a subset $S$ of a field $\mathbb{F}$.
When used as part of rational function interpolation (cf. \sct{sec:int_rat}), all monomials interpolated by Zippel's algorithm are of the same degree $D$ which further reduces the number of non-zero terms $T$ and thus increases the probability of a successful interpolation.

Instead of using a univariate Newton interpolation in Zippel's algorithm for the multivariate polynomial interpolation, \citeres{Kaltofen:2000,Kaltofen:2003} present a racing algorithm which races the sparse Ben-Or/Tiwari algorithm\,\cite{Ben-Or:1988,Kaltofen:1990_2} against the dense Newton interpolation.
This procedure may significantly reduce the number of black-box probes by combining the advantages of both a sparse and a dense algorithm.
Further optimizations of the Ben-Or/Tiwari algorithm are described in \citere{Javadi:2010}.
We leave these optimizations for future versions.

\subsection{Interpolation of rational functions}
\label{sec:int_rat}
Rational functions can be constructed by combining two polynomials. Given two
polynomials $P,Q \in \zp{p}[\vec{z}]$, we define a rational function
$f\in \zp{p}(\vec{z})$, where $\zp{p}(\vec{z})$ is the field of
rational functions in the variables $\vec{z}$, as the ratio of $P$ and $Q$:
\begin{equation}
f(\vec{z}) = \frac{P(\vec{z})}{Q(\vec{z})} = \frac{\sum_{i=1}^{T_\text{n}} n_{\alpha_i} \vec{z}^{\;\alpha_i}}
{\sum_{j=1}^{T_\text{d}} d_{\beta_j} \vec{z}^{\;\beta_j}}.
\end{equation}
The $T_\text{n} \ (T_\text{d})$ non-zero coefficients $n_{\alpha_i} \ (d_{\beta_j})$ are members of the field $\zp{p}$ corresponding to multi-indices $\alpha_i \ (\beta_i)$.

Rational functions are not uniquely defined since their normalization is
arbitrary. In order to provide a unique
representation, we define the lowest degree coefficient in the denominator
to be equal to one. If several monomials contribute to the lowest degree $d_\text{min}$,
we choose to define that
coefficient of the monomial $\vec{z}^{\;\alpha}$ to be equal to one whose
multi-index $\alpha$ is the smallest in a colexicographical ordering, e.g.\
\begin{equation}
(1,1,0) < (1,0,1) < (0,1,1),
\end{equation}
for $d = 2$.

The best strategy for interpolating rational functions is highly dependent on the available information.
In the univariate case one can use Thiele's interpolation formula\,\cite{Abramowitz}.
In the multivariate case, the best algorithm in terms of black-box probes is to solve a dense system of equations if one knows the terms of the rational function but not the values of the coefficients.
Of course, this also works if one can restrict the number of terms by knowing bounds for the degrees of numerator and denominator.
However, this is then a completely dense interpolation, since the number of possible terms is in general larger than the number of non-zero terms $T = T_\text{n} + T_\text{d}$.
In the early 1990s, Grigoriev, Karpinski, and Singer presented an algorithm to obtain bounds and then build a system of equations\,\cite{Grigoriev:1990,Grigoriev:1991,Grigoriev:1994}.
This approach can have a bad scaling behavior for many unknowns and high degrees since Gaussian elimination scales as $O(m^3)$ in time and $O(m^2)$ in space, with $m$ being the number of equations.

A different algorithm was presented by Kaltofen and Trager in 1990\,\cite{Kaltofen:1990_1}.
It evaluates numerator and denominator of a rational function separately, which allows to perform sparse polynomial interpolations.
This algorithm was refined by Kaltofen and Yang in 2007\,\cite{Kaltofen:2007}.
In 2005, de Kleine, Monagan, and Wittkopf published the algorithm described in \citere{deKleine:2005} which is based on a modified version of the Extended Euclidean Algorithm\,\cite{von_zur_Gathen,Monagan:2004}.
Huang and Goa presented an algorithm in 2017\,\cite{Huang:2017} based on Kronecker substitution\,\cite{von_zur_Gathen}.
These three algorithms all require bounds on the degrees of numerator and denominator.

In 2011, Cuyt and Lee presented an algorithm which uses the results of univariate interpolations of a rational function as input for multivariate polynomial interpolations\,\cite{Cuyt:2011}.
It is this algorithm which we adopt for \code{FireFly}, because it requires no information apart from the number of variables a priori and it utilizes the sparsity of a rational function.

Thus, we start by discussing how to interpolate a univariate rational function. According to Thiele\,\cite{Abramowitz} one can express a rational function $f\in \zp{p}(t)$ as
a continued fraction
\begin{equation}
\label{eq:thiele}
\tau(t) = b_0 + (t - t_1) \left(b_1 + (t - t_2)\left(b_2 + (t - t_3)\left(\dots + \frac{t - t_{N}}{b_{N}}\right)^{-1}\right)^{-1}\right)^{-1} ,
\end{equation}
with $t_1,\dots, t_{N+1}$ being distinct elements of $\zp{p}$. The coefficients $b_0\dots, b_N$ can be obtained recursively by
numerical evaluations of the rational function $f$ at $t_1,\dots, t_{N+1}$,
\begin{align}
b_i &\equiv b_{i,i},\\
\label{eq:t_rec_2}
b_{i,j} & = \frac{t_{i + 1} - t_j}{b_{i,j-1} - b_{j-1}},\\
b_{i,0} &= f(t_{i + 1}).
\end{align}
The termination criterion is reached if one finds agreement between $f(t_i)$ and $\tau(t_i)$.
Note that this approach only scales optimally with the number of black-box probes if the degree
of the polynomials of numerator and denominator differ at most by one.

In \eqn{eq:t_rec_2} unlucky zeros can occur in the denominators with a probability bounded by the Zippel-Schwartz lemma.
This leads to the failure of the algorithm.
However, it can be circumvented by interpolating with different $t_i$ or in a different field.
Note that \eqn{eq:thiele} is not the only method for rational function interpolation. One can also utilize the Extended Euclidean Algorithm as shown in \citeres{von_zur_Gathen, Khodadad:2006}.

The idea of the algorithm of \citere{Cuyt:2011} is to perform the multivariate rational function interpolation with a dense
univariate rational function interpolation and a sparse multivariate polynomial interpolation. First, we assume a rational function $f(\vec{z})$ of $n$ variables which has a constant in the denominator and discuss the general case later. The constant is used to normalize the rational function and set to one.
One starts by introducing a homogenization variable $t$ and defining a new function $\tilde{f}(t\vec{z})$ as\,\cite{Diaz:1998}
\begin{equation}
\tilde{f}(t\vec{z}) = f(tz_1,\dots,tz_n).
\end{equation}
We can interpret $\tilde{f}$ as a univariate rational function in the variable $t$, whose coefficients are multivariate polynomials in $\vec{z}$.
Their degrees are bounded by the corresponding degree of $t$.
Thus, by interpolating $\tilde{f}$ in $t$, we can use its coefficients for multivariate polynomial interpolation as described in \sct{ssec:int_poly}.
As optimization, we can set $z_1 = 1$ and reconstruct its power by homogenizing with respect to the corresponding power of $t$.
To interpolate $\tilde{f}$, we generate random anchor points $y_{i,1}$ for the remaining $z_i$ and perform a Thiele interpolation by evaluating $\tilde{f}(t\vec{y}^{\;1})$ for several random $t$ until it terminates.
We proceed for each variable sequentially ($z_1\to z_2 \to \dots\to z_n$) until the polynomial interpolation terminates.
\citere{Cuyt:2011} uses the racing algorithm of \citeres{Kaltofen:2000,Kaltofen:2003} mentioned in \sct{ssec:int_poly} for the polynomial interpolation instead of Zippel's algorithm with a univariate Newton interpolation as we do.

Since Thiele's interpolation is not optimal for general rational functions in terms of the number of black-box probes, we use the knowledge gained during the first interpolation and subsequently only solve systems of equations in the homogenization variable $t$ for all remaining coefficients.
Additionally, we remove already solved coefficients from the system, to further reduce the number of black-box probes.
The univariate system of equations can then easily be constructed by evaluating $\tilde f(t\vec{y})$ for different $t$ chosen randomly from $\zp{p}$ at a fixed $\vec{z} = \vec{y}$ and construct each equation as
\begin{equation}
  \label{eq:univariate_equation}
  \sum_i n_{\text{u},i}(\vec{y}) t^{r_i} - \tilde f(t, \vec{y}) \sum_j d_{\text{u},j}(\vec{y}) t^{r_j} = \tilde f(t, \vec{y}) \sum_{j} d_{\text{s},j}(\vec{y}) t^{r_j} - \sum_{i} n_{\text{s},i}(\vec{y}) t^{r_i},
\end{equation}
where the subscripts s and u denote the solved and unsolved coefficients, respectively.
In this notation the normalizing coefficient is included in the solved subset.
Each of the solved coefficients can be computed by evaluating the corresponding multivariate polynomial at $\vec{z} = \vec{y}$.
When enough equations have been obtained, the system can be solved for the unsolved coefficients.
The time spent solving such univariate systems is usually negligible compared to the evaluation of the black-box function.

Until now, we have assumed that the denominator of $f$ has a constant term. Of course the discussion also holds for a constant term in the numerator.
The generalization of the presented algorithm to arbitrary rational functions leads to a normalization problem.
Assuming no constant term in both numerator and denominator, we cannot set the coefficient of the the lowest degree monomial of either the numerator or denominator of $\tilde{f}$ to one since this coefficient is a polynomial and not a constant.
The algorithm presented in \citere{Cuyt:2011} cures this problem by introducing a variable shift $\vec{s} = (s_1,\dots,s_n)$ such that $f(\vec{z}) \to f(\vec{z} + \vec{s}) \equiv \hat{f}(\vec{z})$ and $\hat{f}$ has a constant term.
A reliable way to fulfill this condition is to choose random but distinct values for all $s_i$.
This shift comes with the caveat that $\hat{f}$ will be a much denser rational function in general.
Especially, aside from accidental cancellations, it will contain all univariate terms in $t$ up to the maximal degrees of numerator and denominator.

For simplicity we will focus on the numerator, noting that the following procedure can be applied completely analogously to the denominator.
Assume that the maximal degree of the numerator of $\hat{f}$ in $t$ is $D$ with its corresponding coefficient $n_D$. Note that $n_D$ will not be affected by the shift, thus, one starts by interpolating only this polynomial.
The numerical values of $n_i$ for $i<D$ are stored for later usage.
Once $n_D$ has been interpolated, we can remove it from the system of equations as before to reduce the number of black-box probes, and additionally subtract the effect of the shift on the lower degree coefficients to preserve their sparsity.
The latter can be achieved by analytically evaluating
\begin{equation}
n_D(\vec{z} + \vec{s}) - n_D(\vec{z}).
\label{eq:shift_subtract}
\end{equation}
It is convenient to save the results of \eqn{eq:shift_subtract} split according to their corresponding degree in $t$. Subsequently, one moves to $n_{D-1}$ and subtracts the term of \eqn{eq:shift_subtract} corresponding to the degree $D-1$ evaluated at the same $y_{i,j}$ as the current
interpolation point. This removes the shift from $n_{D-1}$ and preserves its sparsity in the multivariate polynomial interpolation.

To illustrate the above algorithm, we present the following example which will be interpolated over $\zp{509}$.
Assume we want to interpolate the function
\begin{equation}
  f(\vec{z}) = \frac{3z_1 + 7z_2}{z_1 + z_2 + 4z_1z_2}.
  \label{eq:rat_example}
\end{equation}
We start by choosing the anchor points $\vec{y}^{\;1}$ for $\vec{z}$, and homogenize with $t$:
\begin{equation}
y_{1,1} = 1,\qquad y_{2,1} = 10\quad \Rightarrow \quad\tilde{f}\left(t\vec{y}^{\;1}\right) = \frac{73t}{11t + 40 t^2}.
\end{equation}
Since there is no constant term in the denominator, we perform a variable shift $\vec{s}$ to find a unique normalization:
\begin{equation}
s_1 = 4,\qquad s_2 = 1 \quad \Rightarrow \quad \hat{f}\left(t\vec{y}^{\;1}\right)= \frac{316 + 464t}{1+178t+317t^2}.
\label{eqn:rexp_first}
\end{equation}
The function $\hat{f}\left(t\vec{y}^{\;1}\right)$ in \eqn{eqn:rexp_first}, whose coefficients are already uniquely normalized, is the result of the Thiele interpolation in the first step of the algorithm.
Since we now know the maximal degrees of numerator and denominator, we can proceed in the remaining stages by solving systems of equations instead of using \eqn{eq:thiele} to reduce the number of black-box probes.
The numerical coefficients of $\hat{f}\left(t\vec{y}^{\;1}\right)$ are multivariate polynomials evaluated at $t\vec{y}^{\;1}+ \vec{s}$ and,
thus, are used as the input for the following polynomial interpolation according to \sct{ssec:int_poly}.
We will now focus on the numerator.
The denominator can be interpolated in an analogous way. Assume we have fully interpolated
the highest degree of the numerator as
\begin{equation}
P'_1 = 291 + 170z_2,
\end{equation}
which will yield
\begin{equation}
P_1 = 291z_1 + 170z_2
\end{equation}
after homogenization.
$P'_1$ is the unhomogenized coefficient. To interpolate $n_0$ without the shift, we have to remove its effect before performing the polynomial interpolation.
The input for the latter
will thus be
\begin{equation}
316 - (P_1(\vec{z} + \vec{s}) - P_1(\vec{z})) = 316 - (146 + 170) = 0.
\end{equation}
This is expected since there is no constant term in the numerator of $f(\vec{z})$. In total it takes twelve black-box probes to interpolate \eqn{eq:rat_example} using this algorithm.
One can also apply this example to the case where $d_0\neq 0$ with $\vec{s}=0$. To further reduce the probes mandatory for this algorithm to succeed, the variable with the highest degree should be set to one and is thus only obtained by homogenization of the degrees after the polynomial interpolation finishes. This step requires neither additional black-box evaluations nor the use of interpolation algorithms.

Note that there is no need that the denominator comes with a constant. The same procedure can be applied if one finds a constant in the numerator after introducing a shift and using the latter for normalization. Additionally, one can apply algorithms to find a shift in fewer variables to reduce the higher complexity introduced by the normalization ambiguity. A straightforward algorithm is to test different shifts and apply the one, which shifts a minimal number of variables to obtain a unique normalization. To achieve this, in a first run the maximal degrees of numerator and denominator of the homogenized rational function have to be determined. Since we assume no knowledge about the black box, it is necessary to shift all variables to guarantee the existence of constants in numerator and denominator with high probability and thus prevent cancellations in the homogenization variable. Afterwards, one tests different shift configurations and checks if the maximal degrees of numerator and denominator coincide with the degrees obtained when shifting all variables. These steps are important since a shift including all arguments can lead to $O\left(\binom{n + D}{D}\right)$ additional terms, which have to be calculated to remove the effect of the shift in the interpolation of multivariate polynomials.

The presented approach above, based on \citere{Cuyt:2011}, aims to preserve the sparsity of the rational function. However, for dense functions this algorithm requires more black-box probes than a dense interpolation, where one removes the shift only after the complete interpolation of the rational function.
The reason for this is that in a dense interpolation one can interpolate low degree terms early and remove them from the homogenized system of equations.
This is not possible in the sparse approach because it interpolates the terms starting from the highest degrees.
A hybrid approach where one interpolates lower degree terms densely and high degree terms sparsely might offer the advantages of both methods.
We plan to study this for future versions.

\subsection{From finite fields to rational numbers}
\label{sec:ff_to_q}
In the previous sections, we described how one can interpolate multivariate
polynomials and rational functions over a finite field. Since their
coefficients are only valid in the specific field used for
the interpolation, one has to promote these numbers to rational numbers
such that the corresponding functions are elements of $\mathbb{Q}[\vec{z}]$ and
$\mathbb{Q}(\vec{z})$, respectively.
Therefore, we need to reconstruct the
interpolated results over the rational field. Generally, there is no inversion
of the mapping from rational numbers to members of a finite field, but one can use a
method called rational reconstruction (\abbrev{RR}).
This method is based on the Extended Euclidean Algorithm\,\cite{von_zur_Gathen} and the first algorithm was described by Wang in 1981\,\cite{Wang:1981}, which is presented in \alg{alg:euclid}.

This algorithm leads to a guess for a rational number $a = n/d$ from its image
\begin{equation}
e = a\;\text{mod}\;m,
\end{equation}
where $n$, $d$, and $m > e \geq 0$ are integers.
It will succeed if $|n|,|d| \leq \sqrt{m/2}$.
In \citere{Wang:1982} it was proven that $a$ is unique if it is found for a given $e$ and $m$.
However, this does not mean that the $a$ is the correct rational number in $\mathbb{Q}$ which one desires to reconstruct, because the unique guess can be different for different moduli $m$.
Also, the bound of $\sqrt{m/2}$ could lead to failures of the \abbrev{RR} if $m$ is restricted to machine-size integers.

\begin{algorithm}
\caption{Modified Euclidean Algorithm for rational reconstruction based on \citere{Wang:1981}.}
\label{alg:euclid}
\algrenewcommand\algorithmicrequire{\textbf{Input:}}
\algrenewcommand\algorithmicensure{\textbf{Output:}}
\begin{algorithmic}
\Require {Integers \code{m} $>$ \code{e} $\geq$ 0.}
\Ensure {A rational number \code{n / d} such that \code{n / d} \text{mod} \code{m} = \code{e} or FAIL.}
\Function{rational\_reconstruction}{\code{e}, \code{m}}
\State \code{t} $\gets$ \code{1};\quad \code{old\_t} $\gets$ \code{1};
\State \code{r} $\gets$ \code{m};\quad \code{old\_r} $\gets$ \code{e};
\While{\code{2 * r**2} $>$ \code{m}}
\State \code{quotient} $\gets$ \code{$\lfloor$ old\_r  / r $\rfloor$};
\State \code{(old\_r , r)} $\gets$ \code{(r, old\_r - quotient * r});
\State \code{(old\_t, t)} $\gets$ \code{(t, old\_t - quotient * t)};
\EndWhile
\If{\code{2 * t**2} $>$ \code{m} \textbf{ or } \texttt{gcd}(\code{r}, \code{t}) $\neq$ 1}
\State Throw an error that the rational reconstruction failed;
\EndIf
\State \code{n} $\gets$ \code{r};\quad \code{d} $\gets$ \code{t};
\State\Return \code{sign(d) * n / |d|};
\EndFunction
\end{algorithmic}
\end{algorithm}

Both problems can be solved by the Chinese Remainder Theorem (\abbrev{CRT})\,\cite{von_zur_Gathen}, which
comes with the cost of additional interpolations in other prime fields. The theorem states that
one can uniquely reconstruct an element in $\zp{m}$ from its images in
$\zp{p_i}$ with $i=1,\dots,l$ by combining pairwise coprime numbers
$p_i$ to $m=p_1\cdot p_2\; \cdots \; \cdot p_l$ and applying the \abbrev{CRT}. An algorithmic realization is shown in \alg{alg:cr}.
To reconstruct a rational number, one can then combine as many prime numbers as are required to successfully perform the \abbrev{RR}.
We then consider the rational number to probably be the correct number in $\mathbb{Q}$ if the \abbrev{RR} yields the same result for two different fields.
This guess will be tested later.

\begin{algorithm}
\caption{Chinese Remainder Theorem algorithm \abbrev{CRT} based on \citere{von_zur_Gathen}.}
\label{alg:cr}
\algrenewcommand\algorithmicrequire{\textbf{Input:}}
\algrenewcommand\algorithmicensure{\textbf{Output:}}
\begin{algorithmic}
\Require {Two pairs containing a prime \code{p\_i} of a finite field and a coefficient \code{c\_i}
obtained by the functional reconstruction over this field.}
\Ensure {A pair consisting of the combined value of the \code{c\_i}'s and \code{p\_i}'s according to the \abbrev{CRT}.}
\Function{chinese\_remainder}{(\code{c\_1}, \code{p\_1}), (\code{c\_2}, \code{p\_2})}
\State \code{p\_3} $\gets$ \code{p\_1 * p\_2};
\State {\code{m\_1}} $\gets$ (\code{p\_2}$^{-1}$ {\code{\%}} {\code{p\_1}}) {\code{*}} {\code{p\_2}};
\State \code{m\_2} $\gets$ (\code{1 - m\_1}) \code{\% p\_3};
\State \code{c\_3} $\gets$ (\code{m\_1 * c\_1 + m\_2 * c\_2}) \code{\% p\_3};
\State\Return (\code{c\_3}, \code{p\_3});
\EndFunction
\end{algorithmic}
\end{algorithm}

\alg{alg:euclid} is not an optimal algorithm for arbitrary $n$ and $d$ because it will only succeed if both $|n|$ and $|d|$ are smaller than $\sqrt{m/2}$.
Thus, in the worst case they differ by many orders in magnitude and only one of them fails the bound.
In principle, it is possible to have different bounds for $n$ and $d$.
However, without additional knowledge one does not know a priori how to choose them.
In \citere{Monagan:2004} it was observed that the guess of the rational number comes together with a huge quotient in the Euclidean Algorithm.
The suggested algorithm based on this observation is called Maximal Quotient Rational Reconstruction and presented in \alg{alg:mqrr} in \sct{sec:algorithms}.
It can only be proven that it returns a unique solution if $|n||d| \leq \sqrt{m}/3$.
However, it performs much better in the average case because large quotients from random input are rare.
The parameter $T$ allows to choose a lower bound for the quotients and, thus, the tolerance for false positive reconstructions.
For \code{FireFly} we adjusted it to get roughly $1\,\%$ false positive results.

We then race \alg{alg:euclid} against \alg{alg:mqrr} and consider a guess for a rational number as probably correct if either of the two algorithms reconstructs the same number in two consecutive rings.
Both algorithms can return false positive results, i.e.\ unique rational numbers which are not the correct number in $\mathbb{Q}$.
However, we consider the probability to reconstruct the same false positive number in two different rings as negligible.
We also consider a guess as probably correct if the combined integer does not change after applying the \abbrev{CRT}.

Combining the \abbrev{RR}, the \abbrev{CRT},
and the interpolation of functions over finite fields, the generic algorithm to
obtain an interpolated function in $\mathbb{Q}$ is shown in \alg{alg:full_rec_alg}.
The algorithm succeeds when the function build up by the probably correct coefficients in $\mathbb{Q}$ coincides with the black-box probe evaluated in a new prime field.
We also take all coefficients into account where either \alg{alg:euclid} or \alg{alg:mqrr} was able to reconstruct a guess, which does not fulfill the criteria mentioned above, because it still could be correct.

\begin{algorithm}
\caption{Functional reconstruction algorithm over $\mathbb{Q}$ based on \citere{deKleine:2005}.}
\label{alg:full_rec_alg}
\algrenewcommand\algorithmicrequire{\textbf{Input:}}
\algrenewcommand\algorithmicensure{\textbf{Output:}}
\begin{algorithmic}
\Require {A sequence of primes $p_1,\dots,p_l$ and a black-box function $f(\vec{z}) \equiv f(z_1,z_2,\ldots,z_n)$.}
\Ensure {The result of $f(\vec{z})$ in $\mathbb{Q}$ with high probability.}
\For{$i\gets 1; i\le n$}
\If{$i$$=$1}
\State Interpolate the function $f$ over $\zp{p_1}(\vec{z})$ and store the result;
\State Use the \abbrev{RR} to promote the stored result to a guess $g\in\mathbb{Q}(\vec{z})$;
\Else
\If{the previous rational reconstruction succeeded}
\State Evaluate $g$ over the new field $\zp{p_{i}}$ for a given $\vec{z}$;
\If{$g(\vec{z})$ $=$ $f(\vec{z})$}
\State Terminate the algorithm and set $g(\vec{z}) = f(\vec{z})$;
\EndIf
\Else
\State Interpolate the function $f$ over $\zp{p_i}$ and store the result;
\State Use the \abbrev{CRT} to combine the function in $\zp{p_i}$ with the
stored result in $\zp{p_1\cdots p_{i-1}}$
\State and store the combined result;
\State Use the \abbrev{RR} to promote the stored result in $\zp{p_1\cdots p_i}$ to a guess $g\in\mathbb{Q}(\vec{z})$;
\EndIf
\EndIf
\State $i \gets i + 1;$
\EndFor
\end{algorithmic}
\end{algorithm}

Given that all probabilistic assumptions hold, one obtains full knowledge about the functional form after the first prime field and only the coefficients have
to be promoted to $\mathbb{Q}$. Thus, one can use this knowledge and only reconstruct those coefficients which cannot be promoted
to $\mathbb{Q}$ or are not valid in all used prime fields. This can be done most efficiently in numbers of black-box probes by
solving systems of equations and remove all already known coefficients.
Additionally, no variable shift is required to solve the normalization problem anymore.

However, the simple idea to construct a single system of equations for all coefficients of a rational function which still need to be determined has a major drawback.
It requires $O(T^3)$ operations to solve this system with a standard algorithm.
The size of the system can become quite large if the function has many variables and is densely populated.
Solving this system can then become much more expensive than just interpolating the function from scratch if the time cost of the black-box probes is not too high. For polynomials, however, this can be directly solved by using a generalized transposed Vandermonde system \cite{Zippel:1990}.

To circumvent this problem for rational functions, we again utilize a homogenization variable $t$.
We first build a system of equations in the homogenization variable for all degrees which contain unsolved coefficients by choosing the same ansatz (\ref{eq:univariate_equation}) as for the interpolation, namely, evaluating the black box for randomized $t$ and a fixed value of $\vec{z}$ using anchor points.
Since these systems only contain powers of $t$, they are in general much smaller than the full unhomogenized system, and, thus cheap to solve. The resultant coefficients of orders of $t$ are again multivariate polynomials evaluated at some parameter point in all remaining variables. These coefficients can then be used as input for a generalized and shifted Vandermonde system, i.e., for each degree of $t$ such a system is built.
As soon as we have enough values to solve one of these systems, we solve it and remove the corresponding degree of $t$ from the following homogenized systems of equations.
Therefore, the number of black-box probes is the same as for the full multivariate system of equations for the rational function without homogenization.

However, a rational function has to be normalized to one of the coefficients in order to yield unambiguous results after solving the homogenized system.
Assuming that the functional form is known after the interpolation over the first prime field, we check all degrees for the number of contributing monomials and if their coefficients have already been promoted to $\mathbb{Q}$.
If there is only a single contributing monomial for any degree, we choose its coefficient as normalization by setting it to one. For illustration, let its degree be $d$.
Afterwards, we check again which monomial coefficients can be promoted to $\mathbb{Q}$ using this normalization.
Therefore, the homogenized rational function evaluated at $\vec{z} = \vec{y}$ is given by
\begin{equation}
  \tilde f(t, \vec{y}) = \frac{\vec{y}^{\;\alpha} t^d + \sum_i n_{i}(\vec{y}) t^{d_i}} {\sum_j d_{j}(\vec{y}) t^{d_j}} \qquad \text{or} \qquad \tilde f(t, \vec{y}) = \frac{\sum_i n_{i}(\vec{y}) t^{d_i}} {\vec{y}^{\;\alpha} t^d + \sum_j d_{j}(\vec{y}) t^{d_j}},
\end{equation}
where $n_{i}(\vec{y})$ and $d_{j}(\vec{y})$ are the coefficients of the homogenized rational function evaluated at $\vec{z} = \vec{y}$, which are multivariate polynomials.
The homogenized system can then be built exactly as in \eqn{eq:univariate_equation} and its solutions for $n_{\text{u},i}(\vec{y})$ and $d_{\text{u},j}(\vec{y})$ serve as input for the generalized and shifted Vandermonde systems as described in \sct{sec:int_rat}.

In the application to \abbrev{IBP} reductions as described in \sct{sec:IBP} a normalizing coefficient can be found sometimes.
This is due to the fact that the variables are the space-time dimension $d$ and variables with a mass dimension.
The rational functions occurring have a fixed mass dimension and, thus, numerator and denominator have fixed mass dimensions as well.
Different powers of $d$ then distribute the monomials to different degrees in the homogenizing variable $t$.
The known mass dimensions also make it possible to set one of the mass variables to one, because it can be unambiguously reconstructed after the calculation.
This distributes the monomial degrees further.

The normalization can also be fixed if the coefficient of a degree in $t$ is completely known represented as multivariate polynomial, i.e.\ all coefficients of the polynomial are already promoted to $\mathbb{Q}$.
The numerical value of the coefficient can then always be computed for all $\vec{z}$ and all prime fields, which allows us to use it for normalization.

Should there be no coefficient which can be used for normalization, we reintroduce a shift so that we can normalize to the constant term.
We then proceed as described above with the complication that we have to start with the highest degree and then remove the shift from lower degrees recursively as described in \sct{sec:int_rat}.

Note that there is always a non-vanishing probability that one hits unlucky primes, zeros, or that Zippel's algorithm gets spoiled by bad anchor points. In such cases our presented algorithm fails since it assumes that all these cases do not occur. If these assumptions are not fulfilled, we provide an algorithm which performs new interpolations over additional prime fields with a different set of anchor points instead of using the optimizations described in the previous paragraphs. Therefore, this algorithm is sensitive to ambiguities like unlucky primes, zeros, and bad anchor points. The higher probability of a successful interpolation comes with the cost of a non-negligible amount of additional black-box probes. However, if the black box is proportional to a product of all the used primes, this approach will also fail.

\section{\code{FireFly}}
\label{sec:firefly}
We implemented the functional interpolation and rational number reconstruction algorithms described in the previous section in the \code{C++} library \code{FireFly}, which is publicly available at

\begin{center}
  \href{https://gitlab.com/firefly-library/firefly}{\code{https://gitlab.com/firefly-library/firefly}}
\end{center}

It requires
\begin{itemize}
\item A \code{C++} compiler supporting \code{C++11}
\item \code{CMake} $\ge$ 3.1
\item \code{FLINT} $\ge$ 2.5 (optional)
\item \code{GMP} $\ge$ 6.1
\end{itemize}

The dependency on the \abbrev{GNU} Multiple Precision Arithmetic Library (\abbrev{GMP})\,\cite{gmp} is needed to utilize the \abbrev{CRT} if a single prime field is not sufficient to reconstruct the black-box function.

After downloading the library, \code{FireFly} can be configured and compiled by running
\begin{lstlisting}
cd $FIREFLY_PATH
mkdir build
cd build
cmake -DWITH_FLINT=true .. # Without FLINT: -DWITH_FLINT=false
make
\end{lstlisting}
where \code{\$FIREFLY\_PATH} is the path to the \code{FireFly} directory. When the compilation has finished, the build directory will contain the static and shared libraries \code{libfirefly.a} and \code{libfirefly.so}, respectively.
When compiled without a custom implementation of modular arithmetic (cf.\ \sct{ssec:modular_arithmetic}), the example \code{example.cpp} is compiled as well and the executable can be found in the build directory.
Calling
\begin{lstlisting}[language=C++]
make install
\end{lstlisting}
will additionally copy the corresponding header files and libraries to a specified directory which is by default the system \code{include} and \code{lib} directory, respectively. The prefix of both installation directories can be set with
\begin{lstlisting}
cmake -DCMAKE_INSTALL_PREFIX=$PREFIX ..
\end{lstlisting}
where \code{\$PREFIX} is the desired path prefix. The include and library files will then be installed to \code{\$PREFIX/include}
and \code{\$PREFIX/lib}, respectively.

When using \code{FLINT}\,\cite{flint,Hart:2010} additional flags have to be set if \code{FLINT}s header and library files cannot be found in the system \code{include} and \code{lib} directories. The configuration has thus to be provided with the absolute paths to the include directory and the shared library of \code{FLINT}. This can be done calling
\begin{lstlisting}
cmake -DWITH_FLINT=true -DFLINT_INCLUDE_DIR=$FLINT_INC_PATH -DFLINT_LIBRARY=$FLINT_LIB_PATH ..
\end{lstlisting}
where \code{\$FLINT\_INC\_PATH} and \code{\$FLINT\_LIB\_PATH} are the absolute paths to the include directory and to the shared library of \code{FLINT}, respectively.

In the following sections, we describe the library \code{FireFly} and give examples of its usage. Note that a detailed documentation can be produced using \code{Doxygen} by calling
\begin{lstlisting}
make doc
\end{lstlisting}
in the build directory.

\subsection{Finite field integer and modular arithmetic}
\label{ssec:modular_arithmetic}
Most of the numerical computations are carried out over a finite field modulo a prime $p$. Since there is no built-in \code{C++}
implementation of such an object, we provide the class \code{FFInt} which denotes a finite field integer up to 64-bit. An \code{FFInt} holds basically three \code{uint64\_t} objects: \code{n} which is the element of the field, \code{p} to set the defining prime and \code{p\_inv} which is the inverse limb of \code{p} needed for modular arithmetic when using \code{FLINT} \cite{flint,Hart:2010}.
To enable parallel interpolations of various functions over the same field and to save memory, the latter two are static variables. \code{p\_inv} is private and will be changed automatically if \code{p} changes. All basic arithmetic operations
\begin{equation*}
\code{+},\;\code{-},\; \code{/},\; \code{*},\; \code{+=},\; \code{-=},\; \code{*=},\; \code{/=},
\end{equation*}
relational operators
\begin{equation*}
\code{==},\; \code{!=},\; \code{>},\; \code{<},\; \code{>=},\; \code{<=},
\end{equation*}
and unary operators
\begin{equation*}
\code{+, -, !, ++, --}
\end{equation*}
are implemented. A division by 0 yields 0 as handled by \code{FLINT}. We provide a default implementation of modular arithmetic. Additionally, using the configuration options of the previous section, it is possible to use the \code{FLINT} library or a custom implementation instead. Note that using \code{FLINT} results in much higher performance than our default implementation. For user defined modular arithmetic operators, one has to provide routines for all arithmetic operators and the \code{pow} member function which is mentioned in the next paragraph. Configuring \code{FireFly} with custom operators can be done with
\begin{lstlisting}
cmake -DCUSTOM=true ..
\end{lstlisting}

All public member functions and variables are summarized in the following:

\begin{description}
\item \code{static void set\_new\_prime(uint64\_t prime)}\\ Static function that sets \code{p = prime} and calculates \code{p\_inv}. The latter is only set if one compiles \code{FireFly} with \code{FLINT}.
\item \code{FFInt(const T n)}\\ Creates an \code{FFInt} object by setting its value to $n\;\mathrm{mod}\;p$ where the template \code{T} can be of type \code{FFInt}, \code{mpz\_class}, \code{uint64\_t}, \code{std::string}, or any other primitive integer type.
\item \code{FFInt pow(const FFInt\& e)}\\ Returns $n^{e}\; \mathrm{mod}\; p$.
\item \code{FFInt pow(const FFInt\& a, const FFInt\& e)}\\ Returns $a^{e}\; \mathrm{mod}\; p$.
\item \code{static uint64\_t p}\\ The defining prime $p$ of the field.
\item \code{uint64\_t n}\\ The element $n$ of the prime field $\zp{p}$.
\end{description}

\code{set\_new\_prime} has always to be called first before performing modular arithmetic to define a prime field. In the following code snippet, we show the exemplary usage of the \code{FFInt} class:

\begin{lstlisting}[language=C++]
FFInt::set_new_prime(509);
FFInt a(10);
FFInt b(13);
FFInt c(3);
FFInt d = pow(c, a); // d is now 5
FFInt e = a.pow(b) + c/a; // e is now 355
\end{lstlisting}

All member functions are understood to be part of the \code{firefly} namespace.
\code{FireFly} provides the user with an array of the 100 largest 63-bit primes which can be accessed through the function
\code{primes}. The primes are are stored in a descending order. To get the largest 63-bit prime, one calls $\code{primes()[0]}$ and has to include the header \code{ReconstHelper.hpp}. Note that we are only using \code{C++} \abbrev{STL} containers apart from our custom container classes.

\subsection{Reconstructing functions with \code{FireFly}}
\label{sec:ff_interface}
\code{FireFly} offers the interface class \code{Reconstructor} for the reconstruction of rational functions and polynomials. It is provided with a thread pool\footnote{We adapted the thread pool used in \code{FlexibleSUSY 2.0}\,\cite{Athron:2017fvs}.} to allow for the parallel interpolation of various black-box functions over the same prime field and the promotion of their coefficients to $\mathbb{Q}$. Its member functions can be summarized as:

\begin{description}
\item \code{Reconstructor(uint32\_t n, uint32\_t thr\_n, uint32\_t verbosity = IMPORTANT)}\\ Creates a \code{Reconstructor} object with \code{n} variables and \code{thr\_n} threads. The verbosity levels are \code{SILENT}, \code{IMPORTANT}, and \code{CHATTY}.
\item \code{void enable\_scan()}\\ Enables the scan for a sparse shift at the beginning of the reconstruction. As \code{FireFly} assumes a variable ordering by degree, i.e.\ the first variable has the highest degree, the possible shifts are scanned starting by shifting only the last variable and proceeding variable by variable until the first. After these two variables are scanned starting again with the last two proceeding to the first two etc.
\item \code{void reconstruct()}\\ Performs the reconstruction.
\item \code{vector<RationalFunction> get\_result()}\\ Returns the reconstructed functions as \code{RationalFunction} objects.
\item \code{void set\_tags(const vector<string>\& tags)}\\ Enables storing of intermediate results in the directory \code{./ff\_save}. Each function state is stored under the name given by the corresponding entry in \code{tags}. The same immutable ordering as for the probes in \code{black\_box} is assumed.
\item \code{void set\_tags()}\\ Same as above but set generic tag numbers starting at \code{0}.
\item \code{void resume\_from\_saved\_state(const vector<string>\& file\_paths)}\\ Resumes the reconstruction with the saved files given at the absolute paths \code{file\_paths}.
\item \code{void set\_safe\_interpolation()}\\ Instead of performing the optimizations introduced in \sct{sec:ff_to_q}, \code{FireFly} performs full interpolations over additional prime fields to be sensitive to unlucky primes and zeros. This will require a large amount of additional black-box evaluations.
\end{description}

To perform functional reconstructions, the user has to implement a \code{C++} functor derived from the \code{BlackBoxBase} class. The operator \code{()} and the member function \code{prime\_changed} have thus to be declared, with \code{()} being used for the evaluation of black-box probes. \code{prime\_changed} can be used if some variables of the functor have to be changed when changing $\zp{p}$. The operator \code{()} is provided with a tuple of values at which the black box should be evaluated. Additionally, \code{()} should return a filled vector of \code{FFInt} objects corresponding to the user defined functions which shall be reconstructed. The \code{Reconstructor} class demands an immutable ordering of the result vector to verify the consistency of the reconstruction.

Below we show an example how one could use a black box as a functor. Here, two functions will be interpolated in parallel.
\newpage
\begin{lstlisting}[language=C++]
// Example of how one can use a functor for the Reconstructor interface
class BlackBoxUser : public BlackBoxBase {
public:
  BlackBoxUser(){};
  virtual vector<FFInt> operator()(const vector<FFInt>& values) {
    FFInt fun1_num = values[0] + 2*values[1].pow(4);
    FFInt fun1_den =  1 + 3 * values[0]*values[1];
    FFInt fun1 = fun1_num / fun1_den; // (z1 + 2 z2^4) / (1 + 3 z1z2)
    vector<FFInt> result;
    result.emplace_back(fun1);
    result.emplace_back(gghh(values));
    return result;
  }
  virtual void prime_changed(){
    // nothing needs to be done here
  }
};
\end{lstlisting}

To run the reconstruction for the above defined black-box functor, the \code{Reconstructor} class could be used as follows:

\begin{lstlisting}[language=C++]
BlackBoxUser bb(); // Construct the functor
Reconstructor reconst(4, 4, bb); // Construct the Reconstructor class with
// 4 variables and 4 threads using the black box bb
reconst.enable_scan(); // Enable the scan for a sparse shift
vector<string> tags = {"fun1","gghh"};
reconst.set_tags(tags); // Set tags to save states after each prime field
reconst.reconstruct(); // Reconstruct the functions
vector<RationalFunction> results = reconst.get_result(); // Get results
string fun1 = results[0].to_string({"x","y","z","w"}); // Rewrite fun1 as
// a string
\end{lstlisting}

We first initialize the \code{Reconstructor} with four variables and four threads.
Then, we enable the scan for a sparse shift and set tags to store intermediate results.
\code{reconstruct} first scans the functions to check if a shift in a subset of the four variables is sufficient to always produce a constant term for the normalization and then performs the interpolation, manages the promotion to new prime fields and tries to reconstruct each coefficient over $\mathbb{Q}$. To parallelize the reconstruction, each probe is evaluated in its own thread and each black-box function is reconstructed by a \code{RatReconst} object  which itself performs the reconstruction single threaded. The results are stored as \code{RationalFunction} objects which can be converted to strings by calling its member function \code{to\_string} and providing a vector of strings with the variable names as the argument.

Should the interpolation be aborted it can be resumed with
\begin{lstlisting}[language=C++]
// Give the absolute paths to the intermediate results
vector<string> file_paths = {"ff_save/fun1_4.txt", "ff_save/gghh_4.txt"};
// Enables to resume from a saved state
reconst.resume_from_saved_state(file_paths);
reconst.reconstruct();
\end{lstlisting}

The file names consist of the tag given above and the last prime counter for which the object was saved.

A more elaborate example can be found in \code{example.cpp}.
More information about the parallelization used in \code{FireFly} can be found in \app{sec:ff_internal}.

\subsection{Benchmarks}
\label{ssec:Benchmarks}
In this section we highlight the influence of different options which can strongly affect the time spent on a reconstruction and the corresponding memory consumption.
For that purpose, we define the following benchmark functions:
\begin{align}
\label{eq:b1}
f_1(z_1,\dots,z_{20}) &= \frac{\sum_{i=1}^{20}z_i^{20}}{\sum_{i=1}^5 \left(z_1z_2 + z_3z_4+z_5z_6\right)^i z_{20}^{35}}\;,\\
\label{eq:b2}
f_2(z_1,\dots,z_5) &= \frac{123456789109898799879870980\left(\left(1+\sum_{i=1}^5 z_i\right)^{17} - 1\right)}{z_4 - z_2 + z_1^{10}z_2^{10}z_3^{10}z_4^{10}z_5^{10}}\;,\\
\label{eq:b3}
f_3(z_1,\dots,z_5) &= \frac{123456789109898799879870980\left(\left(1+\sum_{i=1}^5 z_i\right)^{20} - 1\right)}{z_4 - z_2 + z_1^{10}z_2^{10}z_3^{10}z_4^{10}z_5^{10}}\;,\\
\label{eq:b4}
f_4(z_1,\dots,z_5)&= \frac{z_1^{100} + z_2^{200} + z_3^{300}}{z_1 z_2 z_3 z_4 z_5 +z_1^4z_2^4z_3^4z_4^4z_5^4}\;.
\end{align}
We test a sparse function with 20 variables ($f_1$), two dense functions with five variables ($f_2,f_3$) which require the usage of the \abbrev{CRT}, and a sparse function with five variables and high individual degrees ($f_4$).
All functions need a shift in at least one variable.

For the benchmarks we measure the total time, the number of black-box probes, and the total memory consumption using a single threaded setup without a thread pool.
We compare different variable orderings and the influence of the scan for a sparser shift to the default options. All calculations are done on an Intel Core i7-3770 and 8\,GiB of \abbrev{RAM}. \code{FLINT} is enabled for modular arithmetic. The results of the benchmark tests are shown in \tab{tab:benchmarks_ff} using different kinds of possible optimizations. To reconstruct $f_2$ and $f_3$ we use the first four and five entries of the \code{primes} vector, respectively, which can be found in the \code{ReconstHelper.hpp} file.

\begin{table}[h]
  \begin{center}
    \caption[]{\label{tab:benchmarks_ff} Benchmarks for $f_1$, $f_2$, $f_3$, and $f_4$ defined in Eqs. (\ref{eq:b1})-(\ref{eq:b3}) obtained with \code{FireFly} using different optimizations.}
    {\renewcommand{\arraystretch}{1.3}%
    \begin{tabular}{c|c|c|c|c|c}
      \toprule
      Function & Shift scan & Variable order & Probes & Runtime & Memory usage \\
      \midrule
      $f_1$ & \xmark & $(z_1,\dots, z_{20})$ & 87138 & 3.6\,s & 27.9\,MiB \\
      $f_1$ & \xmark & $(z_{20}, z_1,\dots,z_{19})$ & 41628 & 2.1\,s & 25.4\,MiB \\
      $f_1$ & \cmark & $(z_1,\dots, z_{20})$ & 84569 & 4.0\,s & 132.1\,MiB \\
      $f_1$ & \cmark & $(z_{20}, z_1,\dots,z_{19})$ & 22617 & 1.7\,s & 127.3\,MiB \\
      \midrule
      $f_2$ & \xmark & $(z_1,\dots, z_{5})$ & 162683 & 50.5\,s & 54.9\,MiB \\
      $f_2$ & \cmark & $(z_1,\dots, z_{5})$ & 155231 & 42.8\,s & 25.2\,MiB \\
      \midrule
      $f_3$ & \xmark & $(z_1,\dots, z_{5})$ & 332894 & 3\,min 8\,s & 54.9\,MiB \\
      $f_3$ & \cmark & $(z_1,\dots, z_{5})$ & 320801 & 2\,min 40\,s & 43.2\,MiB \\
      \midrule
      $f_4$ & \xmark & $(z_1,\dots, z_{5})$ & 139512 & 42.4\,s & 13.7\,MiB \\
      $f_4$ & \xmark & $(z_3,z_2,z_1,z_4,z_5)$ & 54212 & 9.5\,s & 9.4\,MiB \\
      $f_4$ & \cmark & $(z_1,\dots, z_{5})$ & 137295 & 38.0\,s & 13.5\,MiB \\
      $f_4$ & \cmark & $(z_3,z_2,z_1,z_4,z_5)$ & 34349 & 2.8\,s & 7.7\,MiB \\
      \bottomrule
    \end{tabular}}
  \end{center}
\end{table}

To fully reconstruct $f_1$, \code{FireFly} needs $\sim90000$ black-box probes using a parameter shift in all variables and a non-optimal ordering.
This can be reduced to $\sim20000$ probes by using an optimal ordering and scanning for a sparse shift.
The ordering alone leads to a reduction of $\sim 40000$ probes, the search for a sparse shift requests 184 black-box evaluations but additionally reduces the total number of probes by $\sim 20000$ leading to a reduced runtime of more than $50\%$ from 3.6\,s to 1.7\,s taking both optimizations into account.
Note that the memory consumption drastically increases when performing a shift scan for $f_1$ since we first calculate all possible permutations of the shift and store them in a vector.
For 20 variables, this leads to $O(10^6)$ entries occupying $\sim 120$ MiB of memory.
The memory consumption of the reconstruction itself only needs $\sim 7$ MiB and is thus reduced by roughly a factor of 4.
The generation of the permuted shift vector requires 1.3\,s whereas the reconstruction only takes 0.4\,s.
$f_1$ is constructed in a way that the ordering has a huge impact on the number of black-box probes required to reconstruct it.
Thus, scanning only for a sparse shift and not reordering parameters leads to a shift in $z_{20}$ which is the worst shift one could possibly choose.
Therefore, almost no probes are avoided compared to shifting all variables and using the same variable ordering.

The optimization effects seen for $f_1$ are not that drastic when reconstructing dense functions like $f_2$ and $f_3$ in which a parameter reordering is not helpful.
Instead, only a scan for a sparser shift can avoid some probes and reduces the overall runtime in operations in which the shift needs to be removed. Thus, there is no significant benefit concerning memory consumption but a slightly faster runtime. Separated to different prime fields, the number of requested probes using a shift scan for $f_2$ are: 204 to find a suitable shift, 102358 to interpolate the full functional dependence, 26334 for the second prime field since we can normalize to the univariate degree 50 term in the denominator and disable the shift, 26333 for the third prime field, and an additional probe to check that the reconstruction succeeded.
$f_3$ has approximately twice as much terms as $f_2$ so that we can quantify the scaling of the implemented algorithm. We find that the number of probes required to interpolate $f_3$ scales linearly compared to $f_2$.

As for the reconstruction of $f_1$, one can observe similar benefits utilizing optimizations for $f_4$.
A reordering of variables can significantly reduce the number of black-box probes to approximately one third of the non-optimal ordered case and, additionally, shorten the runtime to almost one fifth.
Scanning for a sparser shift without a reordering leads, by construction, only to a small number of black-box evaluations avoided compared to shifting all variables.
The benefit of this option is reflected in the runtime which reduces by almost 4\,s due to the much simpler polynomials needed to remove the effects of the shift during the polynomial interpolation.
Combining an optimal variable order and a sparse shift can further reduce the runtime to 2.8\,s while requiring only less than a quarter of the black-box probes needed using no optimizations.

Usually, there is no knowledge about the function to be interpolated.
Therefore, a useful a priori assumption about the variable ordering cannot be made in general. However, for physical calculations it can be often useful to first study a similar but simpler problem to estimate the variable structure and apply a corresponding ordering to the actual calculation. It is always better to set the potentially highest degree variable to $z_1$ which will thus not be interpolated directly.

\section{Application to \abbrev{IBP} reductions}
\label{sec:IBP}
In multi-loop calculations in high-energy physics one ends up with scalar Feynman integrals after an appropriate tensor reduction.
The general form of scalar Feynman integrals with $L$ loops is given by
\begin{equation}
  \label{eq:scalar_int}
  I(d, \{p_j\}, \{m_i\}, \{a_i\}) \equiv \int_{k_1,\dots ,k_L} \frac{1}{P_1^{a_1}\dots P_N^{a_N}}
\end{equation}
with
\begin{equation}
  \int_k \equiv \int \frac{\mathrm{d}^d k}{(2\pi)^d}.
\end{equation}
and the inverse propagators $P_i = q_i^2 - m_i^2 + \mathrm{i}\epsilon$ in Minkowski space or $P_i = q_i^2 + m_i^2$ in Euclidean space.
The $q_i$ are linear combinations of the loop momenta $k_l$ and the external momenta $p_j$.
The integral $I(d, \{p_j\}, \{m_i\}, \{a_i\})$ depends on the space-time dimension $d$, the set of masses $\{m_i\}$, the set of external momenta $\{p_j\}$, and the propagator powers $a_i$ which take integer values.
$N$ can be computed from $L$ and the number of external momenta $E$ by
\begin{equation}
  N = EL + \frac{L(L+1)}{2}.
\end{equation}

It is useful to define the sum of all positive powers of the propagators of an integral as
\begin{equation}
  r \equiv \sum_{i=1}^{\#\text{prop.}} \theta \left(a_i -\frac{1}{2}\right) a_i
\end{equation}
and the absolute value of the sum of all negative powers as
\begin{equation}
  s \equiv \sum_{i=1}^{\#\text{prop.}} \theta \left(\frac{1}{2}- a_i\right) |a_i|,
  \label{eq:def_s}
\end{equation}
where $\theta(x)$ is the Heaviside step function.
Usually, an integral with higher $r$ or higher $s$ is regarded more difficult than an integral with lower $r$ or $s$.
Therefore, $r$ and $s$ can be used to sort the occurring integrals by difficulty.

The integration-by-parts (\abbrev{IBP}) algorithm of Chetyrkin and Tkachov\,\cite{Tkachov:1981wb,Chetyrkin:1981qh} is based on the observation that inserting the scalar product of a derivative with respect to a loop momentum with another momentum into \eqn{eq:scalar_int} leads to a vanishing integral in dimensional regularization:
\begin{equation}
  \int_{k_1,\dots k_L} \frac{\partial}{\partial k_i^\mu} \left( \tilde q_j^\mu \frac{1}{P_1^{a_1}\dots P_N^{a_N}} \right) = 0,
\end{equation}
where $\tilde q_j^\mu$ can either be another loop momentum or an external momentum.
By explicitly evaluating the derivative one arrives at the linear relations
\begin{equation}
  \label{eq:IBP}
  0 = \sum_n c_n I(d, \{p_j\}, \{m_i\}, \{a_i^{(n)}\})
\end{equation}
with modified $a_i^{(n)}$, where the values change by the addition or subtraction of small integers.
The coefficients $c_n$ are polynomials in $d$, $\{m_i\}$, and $\{p_j\}$ with a small degree and also depend on the $a_i$ in general.
These relations are called \abbrev{IBP} relations.

They can be combined to recursion relations which express an integral through easier integrals.
The recursive application of the relations then allows to reduce all integrals to a small set of master integrals, which have to be computed by other methods to get the final result.
At three-loop level, this approach has been successfully applied for massless propagator-type diagrams\,\cite{Chetyrkin:1981qh,Gorishnii:1989gt,Larin:1991fz}, massive tadpoles\,\cite{Broadhurst:1991fi,Steinhauser:2000ry}, and on-shell propagators\,\cite{Melnikov:1999xp,Melnikov:2000qh,Melnikov:2000zc}.
However, the recursion relations usually have to be derived manually which makes this procedure unfeasible for multi-loop calculations.
On the other hand, the reduction is straightforward and fast should they be found.
In the last decade, some progress in the automatic derivation of recursion relations has been made\,\cite{Lee:2012cn,Smirnov:2013dia,Lee:2013mka,Ruijl:2017cx}.

In 2001, Laporta presented a different strategy\,\cite{Laporta:2001dd}.
By inserting integer values for the $a_i$ in \eqn{eq:IBP}, so-called \textit{seeds}, one obtains a system of equations for the occurring integrals.
These systems can be solved by standard procedures to yield the desired reduction to master integrals.
Obviously, one has to choose a sufficient amount of seeds for the specific problem.
There exist several public implementations of modified versions of the Laporta algorithm, \code{AIR}\,\cite{Anastasiou:2004vj}, \code{FIRE}\,\cite{Smirnov:2008iw,Smirnov:2013dia,Smirnov:2014hma,Smirnov:2019qkx}, \code{Reduze}\,\cite{Studerus:2009ye,vonManteuffel:2012np}, and \code{Kira}\,\cite{Maierhoefer:2017hyi,Maierhofer:2018gpa}, and numerous private ones.
However, the Laporta algorithm has some major drawbacks.
The systems of equations for the state-of-the-art calculations become huge and expensive to solve both in terms of memory and runtime.
This is partially related to large intermediate expressions which occur during the solution process.

Since the \abbrev{IBP} relations are linear, the solution strategies only involve the addition, subtraction, multiplication, and division of the polynomial coefficients $c_n$.
Therefore, the coefficients of the master integrals are rational functions in $d$, $\{m_i\}$, and $\{p_j\}$.
Thus, the problems of the Laporta algorithm can be eased by finite-field techniques as proposed by Kauers in 2008\,\cite{Kauers:2008zz}.
One can replace all occurring variables by members of the finite field and solve the system of equations numerically.
This is in general orders of magnitude faster than solving the system analytically.
The first realized application was to solve the system of equations over a finite field before the actual analytic reduction\,\cite{Kant:2013vta}.
This allows to identify and remove the linearly dependent equations.
By also performing the back substitution, one can additionally identify the master integrals and also select only those equations which suffice to reduce a requested subset of integrals.
This procedure has been implemented in \code{Kira}\,\cite{Maierhoefer:2017hyi}.

Some of the finite-field-interpolation techniques from computer science described in \sct{sec:functional_interpolation} have been summarized before in a physical context in \citeres{vonManteuffel:2014ixa,Peraro:2016wsq}.
The first pure finite-field \abbrev{IBP} reduction was presented by von Manteuffel and Schabinger in 2016\,\cite{vonManteuffel:2016xki}.
Recently, three more calculations were concluded\,\cite{Henn:2019rmi,vonManteuffel:2019wbj,vonManteuffel:2019gpr}.
All of them are one scale problems and, thus, one variable problems, i.e.\ they only require univariate interpolation techniques, which has been implemented in private codes.
The recently published Version 1.2 of \code{Kira} uses a multivariate polynomial interpolation at intermediate stages of the reduction if it seems useful\,\cite{Maierhofer:2018gpa}.
This means that instead of multiplying some (sub-)coefficients algebraically, their numerical values are multiplied and the result is interpolated by Newton's interpolation formula.
However, this interpolation is done over $\mathbb{Z}$ and not in a finite field.

Recently, \code{FIRE6} was published as first public implementation of the Laporta algorithm using the interpolation of rational functions over a finite field\,\cite{Smirnov:2019qkx}.
For polynomials it performs a multivariate Newton interpolation.
The interpolation of rational functions works in several steps.
First, univariate interpolations are performed for each variable separately using Thiele's interpolation formula.
By assuming that the denominator of the final result can be factorized into functions depending only on one variable each, one can turn the rational function into a polynomial by multiplying it with the denominators obtained from the univariate interpolation.
This allows to perform a multivariate Newton interpolation.
However, it seems to only work for up to two variables at the moment and it is not clear whether the factorization assumption can be generalized to several kinematic variables.

For more details on \abbrev{IBP} reductions we refer the reader to the \citere{Grozin:2011mt} and \citeres{Smirnov,Kotikov:2018wxe} for a general overview about techniques for the calculation of multi-loop Feynman integrals.

\subsection{Implementation}
\label{ssec:Kira-Implementation}
Since \code{Kira} already utilizes the reduction over a finite field as a preliminary step, we chose it as a basis for our implementation.
The idea is to use the built-in solver \code{pyRed} to solve the system of \abbrev{IBP}s many times over several prime fields, extract the numerical values, and feed them to \code{FireFly} to reconstruct the rational functions.
We thus added the option to start the reduction with an interpolation in finite fields using \code{FireFly}.
Another important option is to choose the order of the variables of the problem including $d$.
The order can have a major impact on the runtime as shown in \sct{ssec:Benchmarks}.

\makebox[\linewidth][s]{Our implementation uses the \code{Reconstructor} class described in \sct{sec:ff_interface}.
\code{FireFly} is}
\makebox[\linewidth][s]{compiled to use \code{FLINT} for the modular arithmetic.
The black-box functor derived from}
\code{BlackBoxBase} replaces the variables by the values requested by the reconstruction objects and solves the resulting numerical system with \code{pyRed}.
Additionally, we perform two integral selections.
Even though the system only contains all required linearly-independent equations from the start, after the forward elimination not all of these equations are required anymore for the back substitution in general.
We thus select only those equations which are required for the reduction of the requested integrals.
The same selection is performed in the analytic calculation with \code{Kira}.
We perform a second selection after the back substitution which selects only the coefficients of the master integrals for the requested integrals.
This selection does not offer any advantage in an analytic approach because everything is already solved at this point.
In our approach this allows us to omit potentially difficult rational functions which are not required to get the desired result.

\subsection{Examples}
\label{ssec:Kira-Benchmarks}
To illustrate the validity of the reconstruction approach for Laporta's algorithm for multi-scale Feynman integrals, we apply our private implementation described in the previous section to some example topologies.
We compare it with the highly optimized algebraic program \code{Kira 1.2}.\footnote{We use it in combination with version 6.21 of \code{Fermat}\,\cite{fermat}.}
It is important to stress that our link of \code{FireFly} to \code{Kira} has room for optimizations and that we leave real benchmarks for an official implementation.
Therefore, we just compare the straight on calculations, i.e.\ we do not use some of the advanced features of \code{Kira} like the sectorwise forward elimination or the back substitution for subsets of master integrals.
We also do not use the option \texttt{algebraic\_reconstruction}, i.e.\ we compare a fully algebraic calculation to the finite field approach.
In both approaches, we restrict $r$ and $s$ of the seed integrals by $r_\text{max}$ and $s_\text{max}$ and select the integrals with the option \code{select\_mandatory\_recursively} in the same range.

\begin{figure}[h]
\centering
\includegraphics{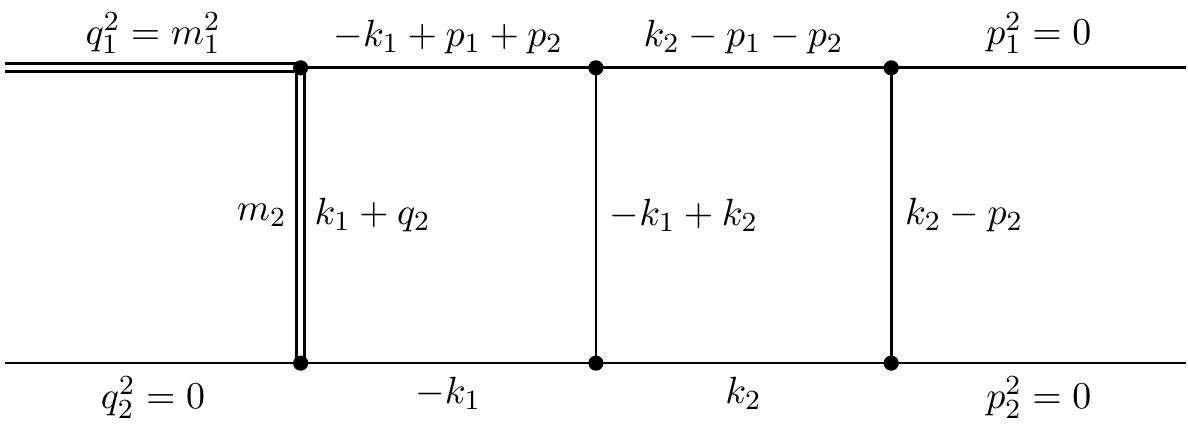}
\caption{The planar double box \code{topo7} which occurs, e.g., in single top production.}
\label{fig:topo7}
\end{figure}

Our first example is \code{topo7}, which is one of the examples provided by \code{Kira}.
This planar two-loop diagram is shown in \fig{fig:topo7}.
It occurs in single top production.
We first set the masses $m_1$ and $m_2$ to zero so that the kinematic invariants $s$ and $t$ are the only massive parameters.\footnote{Not to be confused with the sum of all negative powers $s$ defined in \eqn{eq:def_s}.}
By setting $s$ to one, only $t$ and the space-time dimension $d$ remain as parameters.
For our reconstruction, we order them as $d$, $t$ for the best performance, because $d$ appears with higher powers in the result.

\begin{table}[h]
  \begin{center}
    \caption{Runtime and memory usage for \code{Kira 1.2} and \code{Kira} with \code{FireFly} for \code{topo7} without masses and with $r_\text{max} = 7$.
    We use the option \code{select\_mandatory\_recursively}, order the variables as $d$, $s$, $t$, and set $s$ to one.
    The runtime only includes the time spent on the forward elimination and back substitution or the time for the reconstruction, respectively.
    These tests were run on a computer with an Intel Core i7-3770 and 8\,GiB of \abbrev{RAM}.}
    \label{tab:topo7massless}
    {\renewcommand{\arraystretch}{1.3}
    \begin{tabular}{c|c c|c c c}
      \toprule
      & \multicolumn{2}{c|}{\code{Kira 1.2}} & \multicolumn{3}{c}{\code{Kira} with \code{FireFly}} \\
      $s_\text{max}$ & Runtime & Memory usage & Runtime & CPU time for \code{pyRed} & Memory usage \\
      \midrule
      2 & 7.9\,s & 0.3\,GiB & 1.9\,s & 89\,\% & 0.3\,GiB \\
      3 & 10.2\,s & 0.5\,GiB & 5.2\,s & 90\,\% & 0.5\,GiB \\
      4 & 14.1\,s & 0.9\,GiB & 14.3\,s & 90\,\% & 0.9\,GiB \\
      \bottomrule
    \end{tabular}}
  \end{center}
\end{table}

The results for different $s_\text{max}$ are shown in \tab{tab:topo7massless}.
For $s_\text{max} = 2$, the reconstruction is faster by a factor of four.
However, it scales worse than the algebraic calculation when increasing $s_\text{max}$ with only eight threads available.
\code{Kira} becomes slightly faster for the amplitude with $s_\text{max} = 4$.
\code{Kira} has to reduce 2774, 7994, and 17548 equations for $s_\text{max} = 2,3,4$, whereas the reconstruction implementation selects only the 730, 1837, and 3970 mandatory equations, respectively.
Most of the \abbrev{CPU} time for the reconstruction is used for the numerical solutions of the system with \code{pyRed}, i.e.\ the black-box probes.
For $s_\text{max} = 2$, the reconstruction requires about 120 black-box probes which take 0.07\,s, 180 with 0.17\,s each for $s_\text{max} = 3$, and 280 with 0.33\,s each for $s_\text{max} = 4$.
The forward elimination contributes with roughly 75\,\% to these times.
The memory consumption for both versions is dominated by the preliminary steps to construct the system of linearly independent equations.

The calculation becomes much more expensive in terms of \abbrev{CPU} time when taking the masses into account.
We again order the variables by the highest powers appearing in the result, which now is $m_2$, $t$, $s$, $d$, $m_1$ and set the highest massive parameter $m_2$ to one.

\begin{table}[h]
  \begin{center}
    \caption{Runtime and memory usage for \code{Kira 1.2} and \code{Kira} with \code{FireFly} for \code{topo7} with $r_\text{max} = 7$.
    We use the option \code{select\_mandatory\_recursively}, order the variables as $m_2$, $t$, $s$, $d$, $m_1$, and set $m_2$ to one.
    The runtime only includes the time spent on the forward elimination and back substitution or the time for the reconstruction, respectively.
    These tests were run on a computer with an Intel Core i7-3770 and 8\,GiB of \abbrev{RAM}.}
    \label{tab:topo7}
    {\renewcommand{\arraystretch}{1.3}
    \begin{tabular}{c|c c|c c c}
      \toprule
      & \multicolumn{2}{c|}{\code{Kira 1.2}} & \multicolumn{3}{c}{\code{Kira} with \code{FireFly}} \\
      $s_\text{max}$ & Runtime & Memory usage & Runtime & CPU time for \code{pyRed} & Memory usage \\
      \midrule
      2 & 38\,s & 0.5\,GiB & 128\,s & 94\,\% & 0.4\,GiB \\
      3 & 270\,s & 0.8\,GiB & 880\,s & 91\,\% & 0.8\,GiB \\
      4 & 3000\,s & 1.6\,GiB & 9200\,s & 89\,\% & 3.6\,GiB \\
      \bottomrule
    \end{tabular}}
  \end{center}
\end{table}

The two additional parameters lead to a drastic increase in the runtime up to two orders of magnitude for both approaches for higher $s_\text{max}$, but the reconstruction approach suffers much more than the algebraic one as shown in \tab{tab:topo7}.
However, now the scaling with $s_\text{max}$ is inverted.
The reconstruction approach is slower by a factor of 3.4 for $s_\text{max} = 2$, which reduces to a factor of 3.0 for $s_\text{max} = 4$.
The memory consumption is almost entirely dominated by the actual reduction or reconstruction with $s_\text{max} = 2$ for the reconstruction being the sole exception.
\code{Kira} reduces 3902, 10216, and 22571 equations for $s_\text{max} = 2,3,4$, whereas our reconstruction implementation selects only the 949, 2439, and 5370 mandatory equations.
The number of coefficients which have to be reconstructed increases from 2984 to 7428 to 15531.
Since the complexity of the coefficients also increases when going to higher $s_\text{max}$, they require a lot more memory to store input values, terms, and subtraction terms of the shift.
The higher complexity also leads to a decreasing limitation through \code{pyRed}, because internal calculations in \code{FireFly} become more expensive.
However, the numerical solutions with \code{pyRed} are still the main bottleneck by an order of magnitude.
Roughly 9400 of them are required for $s_\text{max} = 2$, with each of them taking 0.1\,s.
These numbers increase to 25900 and 0.24\,s for $s_\text{max} = 3$ and 117000 and 0.54\,s for $s_\text{max} = 4$.
Again, the forward elimination is responsible for more than 75\,\% of the solution times.

\begin{table}[h]
  \begin{center}
    \caption{Scaling of \code{Kira 1.2} and \code{Kira} with \code{FireFly} with the number of threads for \code{topo7} with $r_\text{max} = 7$ and $s_\text{max} = 4$.
    We use the option \code{select\_mandatory\_recursively}, order the variables as $m_2$, $t$, $s$, $d$, $m_1$, and set $m_2$ to one.
    The runtime only includes the time spent on the forward elimination and back substitution or the time for the reconstruction, respectively.
    These tests were run on a computer with two Intel Xeon Gold 6138 and 768\,GiB of \abbrev{RAM}.}
    \label{tab:topo7_threads}
    {\renewcommand{\arraystretch}{1.3}
    \begin{tabular}{c|c c|c c c}
      \toprule
      & \multicolumn{2}{c|}{\code{Kira 1.2}} & \multicolumn{2}{c}{\code{Kira} with \code{FireFly}} \\
      \# Threads & Runtime & Memory usage & Runtime & Memory usage \\
      \midrule
      10 & 1900\,s & 1.9\,GiB & 7450\,s & 3.6\,GiB \\
      20 & 1500\,s & 2.8\,GiB & 4350\,s & 3.9\,GiB \\
      40 & 1550\,s & 4.8\,GiB & 2950\,s & 4.1\,GiB \\
      80 & 1450\,s & 8.8\,GiB & 2050\,s & 4.9\,GiB \\
      \bottomrule
    \end{tabular}}
  \end{center}
\end{table}

In \tab{tab:topo7_threads} we show how both approaches scale with the number of \abbrev{CPU} threads available.
Note that these numbers are not directly comparable to the numbers in \tab{tab:topo7}, because different computers were used.
The \abbrev{CPU}s not only differ in clock speed and architecture but also in the number of cores.
The Intel Core i7-3770 used in \tab{tab:topo7} has four physical cores and four additional logical cores whereas the two Intel Xeon Gold 6138 used for \tab{tab:topo7_threads} offer 20 physical and 20 additional logical cores each.
Therefore, the calculations in \tab{tab:topo7_threads} can run on physical cores for up to 40 threads.

\code{Kira} does not really profit from many threads in this example.
Doubling the number from 10 to 20 only increases the performance by less than a factor of 1.3.
A further increase basically does not impact the runtime anymore, since large parts of the back substitution run on one or two threads.
However, the memory consumption increases drastically, because each additional \code{Fermat} instance requires additional memory.
On the other hand, doubling the number of threads from 10 to 20 increases the performance of the reconstruction approach by a factor of 1.7, doubling from 20 to 40 by a factor of 1.5, and doubling from 40 to 80 still by a factor of 1.4.
Thus, while the reconstruction takes almost four times longer than the algebraic approach on 10 threads, this difference reduces to only 40\,\% on 80 threads.
Additionally, the reconstruction approach could profit from even more threads.
The performance gain only comes with a moderate increase in memory consumption.

\begin{figure}[h]
\centering\includegraphics{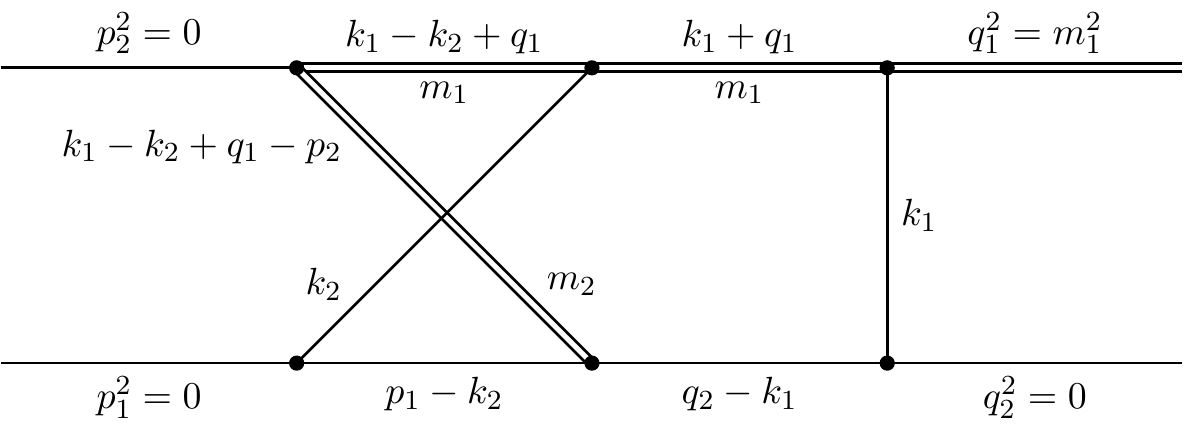}
\caption{The non-planar double box \code{topo5} which occurs, e.g., in single top production.}
\label{fig:topo5}
\end{figure}

Our second example is \code{topo5} of the \code{Kira} examples, which is a two-loop non-planar diagram occurring in single top production.
The diagram is shown in \fig{fig:topo5}.
We order the parameters as $m_1$, $s$, $m_2$, $t$, $d$ and set $m_1$ to one.

\begin{table}[h]
  \begin{center}
    \caption{Runtime and memory usage for \code{Kira 1.2} and \code{Kira} with \code{FireFly} for \code{topo5} with $r_\text{max} = 7$.
    We use the option \code{select\_mandatory\_recursively}, order the variables as $m_1$, $s$, $m_2$, $t$, $d$, and set $m_1$ to one.
    The runtime only includes the time spent on the forward elimination and back substitution or the time for the reconstruction, respectively.
    These tests were run on a computer with two Intel Xeon Gold 6138 and 768\,GiB of \abbrev{RAM}.}
    \label{tab:topo5}
    {\renewcommand{\arraystretch}{1.3}
    \begin{tabular}{c|c c|c c c}
      \toprule
      & \multicolumn{2}{c|}{\code{Kira 1.2}} & \multicolumn{3}{c}{\code{Kira} with \code{FireFly}} \\
      $s_\text{max}$ & Runtime & Memory usage & Runtime & CPU time for \code{pyRed} & Memory usage \\
      \midrule
      1 & 4\,min & 7.6\,GiB & 3\,min & 99\,\% & 0.9\,GiB \\
      2 & 1\,h 53\,min & 33\,GiB & 1\,h 42\,min & 97\,\% & 3.3\,GiB \\
      3 & 18\,h 28\,min & 102\,GiB & 18\,h 34\,min & 91\,\% & 18\,GiB \\
      \bottomrule
    \end{tabular}}
  \end{center}
\end{table}

Both versions perform the reductions in similar times as shown in \tab{tab:topo5}.
However, the reconstruction approach requires significantly less memory.
Again, the CPU time is mainly used for \code{pyRed}.
For $s_\text{max} = 1,2,3$, the reconstruction requires 15500 probes and one prime field, 269700 probes and two prime fields, 1090400 probes and three prime fields.
Each probe takes 0.36\,s, 0.91\,s, or 2.1\,s, respectively, and around 83\,\% of this is spent for the forward elimination.
\code{Kira} reduces 3044, 10069, and 23670 equations whereas the reconstruction approach selects only 434, 1512, and 4086 mandatory equations.

\begin{figure}[h]
\centering\includegraphics{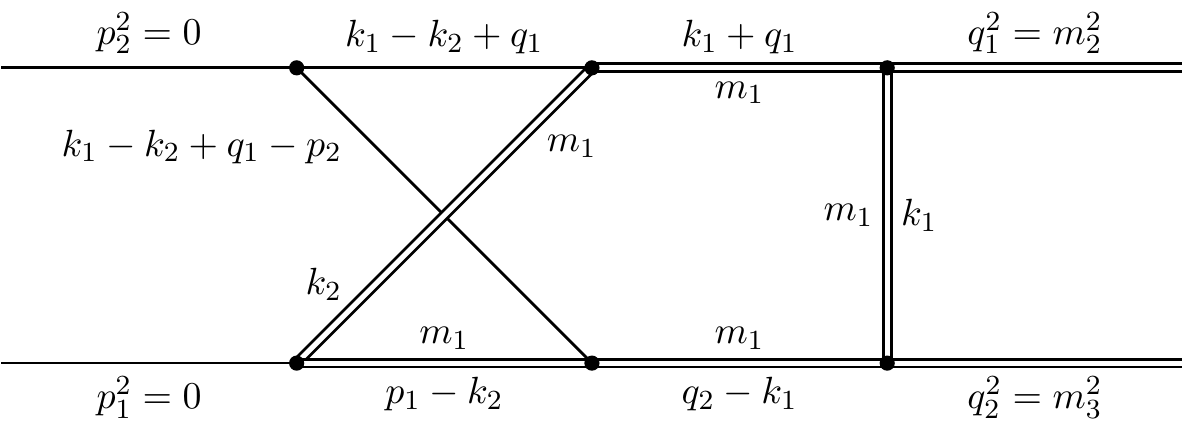}
\caption{The non-planar double box \code{topo5\_m} which occurs, e.g., in the gluon-induced production of a Higgs and a $Z$ boson.}
\label{fig:topo5_m}
\end{figure}

As last example, we consider \code{topo5} with a different mass configuration.
The non-planar two-loop diagram shown in \fig{fig:topo5_m} occurs for example in $HZ$ production through gluon fusion.
In contrast to \code{topo5} there are now five massive parameters: $m_1$, $s$, $m_2$, $m_3$, and $t$.
We can only guess the optimal order from the completed calculations with small $s_\text{max}$.
There, their highest powers occur in the order stated above, with the additional space-time dimension $d$ at the last position.
Thus, we set $m_1$ to one.

\begin{table}[h]
  \begin{center}
    \caption{Runtime and memory usage for \code{Kira 1.2} and \code{Kira} with \code{FireFly} for \code{topo5\_m} with $r_\text{max} = 7$.
    We use the option \code{select\_mandatory\_recursively}, order the variables as $m_1$, $s$, $m_2$, $m_3$, $t$, $d$, and set $m_1$ to one.
    The runtime only includes the time spent on the forward elimination and back substitution or the time for the reconstruction, respectively.
    These tests were run on a computer with two Intel Xeon Gold 6138 and 768\,GiB of \abbrev{RAM}.}
    \label{tab:topo5_m}
    {\renewcommand{\arraystretch}{1.3}
    \begin{tabular}{c|c c|c c c}
      \toprule
      & \multicolumn{2}{c|}{\code{Kira 1.2}} & \multicolumn{3}{c}{\code{Kira} with \code{FireFly}} \\
      $s_\text{max}$ & Runtime & Memory usage & Runtime & CPU time for \code{pyRed} & Memory usage \\
      \midrule
      1 & 22\,h 40\,min & 175\,GiB & 7\,h 35\,min & 88\,\% & 4.7\,GiB \\
      \bottomrule
    \end{tabular}}
  \end{center}
\end{table}

Since this example is extremely complicated, we only consider $s_\text{max} = 1$ in \tab{tab:topo5_m} to illustrate how both approaches perform.
The reconstruction over finite fields is faster by a factor of four and still mainly limited by \code{pyRed}.
Additionally, it requires one to two orders less memory than the algebraic approach.
It selects only the 1785 mandatory equations in contrast to the 3140 equations \code{Kira} reduces.
About $1.5\cdot 10^6$ black-box probes and two prime fields are needed to reconstruct all coefficients.
Each \code{pyRed} solution takes 0.32\,s, where the forward elimination contributes with 93\,\%.

\subsection{Application to the perturbative gradient flow}
The gradient-flow formalism\,\cite{Luscher:2010iy,Luscher:2011bx,Luscher:2013cpa} has proven to be a useful tool in lattice \abbrev{QCD}.
Additionally, there are several areas of potential cross fertilization between lattice and perturbative calculations.
However, the perturbative side was only computed to a low order in perturbation theory in many cases.
We want to use the \abbrev{IBP} reduction as a systematic approach to higher-order calculations.
As a verification of this approach we want to compute the same observable as in \citere{Harlander:2016vzb}, where the computation was performed by solving all integrals numerically.
Furthermore, we want to compute two additional observables.
Our results are published in \citere{Artz:2019bpr}.

Our setup produces all Feynman diagrams, inserts the Feynman rules, and reduces the tensor structures.
We also expand the integrals to the leading order in the quark masses and end up with massless three-loop vacuum integrals of the form
\begin{equation}
  \begin{split}
    &I(d,t,\{t_{f}^\text{up}\}, \{T_i\}, \{a_i\}) = \left(\prod_{f=1}^F\int_0^{t_{f}^\text{up}}\dd t_f\right)\int_{k_1,k_2,k_3} \frac{\exp[-(T_1 q_1^2 + \dots + T_6 q_6^2)]}{q_1^{2a_1} \cdots q_6^{2a_6}}\,,
    \label{eq:intrep}
  \end{split}
\end{equation}
where $F\leq 4$ is the number of flow-time integrations and the upper limits $t_f^\text{up}$ are either the flow-time scale $t$ of the diagram or other flow-time variables $t_{f^\prime}$.
The $T_i$ are nonnegative linear combinations of the flow variables, i.e.\ $T_1 = t + t_1 - 2t_3$.
Due to the expansion in the quark masses, the flow-time scale $t$ is the only massive scale appearing in the integrals.

Instead of evaluating the integrals numerically as in \citere{Harlander:2016vzb} we build a system of equations using \abbrev{IBP}-like relations, where we set $t=1$ and, thus, the space-time dimension $d$ is the only parameter.
However, the algebraic forward elimination with \code{Kira 1.2} fills all our available memory of 768\,GiB of RAM after running for five days and nine hours on the 80 threads of two Intel Xeon Gold 6138.
The progress so far indicates that it probably would have to run for at least several days more if we had enough memory.
It might be possible to complete the reduction by adapting the sectorwise forward elimination introduced in \code{Kira 1.2} for the gradient flow integrals.

Our implementation with finite-field-interpolation techniques as described in \sct{ssec:Kira-Implementation} runs about two days and 20 hours on ten threads to obtain the complete analytic result.
Each numerical system requires around 70\,GiB of RAM and takes $\sim 50$\,s to parse all analytic coefficients to numerical values, $\sim 10400$\,s to complete the forward elimination, and $\sim 16$\,s to complete the back substitution.
We need three 63-bit prime numbers and 201 numerical solutions in total to reconstruct all rational functions with all rational coefficients and an additional solution with a fourth prime to verify the results.

The highest degree in the numerator is 32 and the highest degree in the denominator 34.
At intermediate stages of the analytic forward elimination monomials with at least degree 508 occur.\footnote{We know this because we had to change the \code{Fermat}\,\cite{fermat} settings to support higher degrees with \code{Kira 1.1}.}
Therefore, this is either a prime example for circumventing large intermediate expressions or of the possibility to neglect large coefficients which are not required for the desired result.
However, this is a special problem with only one massive scale, which can be set to one, and, thus, only the space-time dimension $d$ appears in the rational functions.
The reduction is also heavily limited by the forward elimination.
While this is the default case for numerical solutions, this also holds for the analytic reduction of gradient-flow integrals based on our experience.
This is unusual for the reduction of Feynman integrals.

The reduction allows us to express the 3195 required integrals through 188 master integrals.
We checked that our reduction yields the correct results by comparing the numerical values for the integrals expressed through the master integrals with the numerical values of the integrals computed directly.

\section{Conclusions}
\label{sec:conlusions}
We presented the open-source \code{C++} library \code{FireFly} for the reconstruction of polynomials and rational functions, which is based on algorithms developed in computer science.
Additionally, we made some modifications to reduce the number of black-box probes and avoid cancellations.
We showed that \code{FireFly} is capable to reconstruct multivariate rational functions with many variables.

As an example, we applied \code{FireFly} to the Laporta algorithm for \abbrev{IBP} reductions by linking it to \code{Kira}.
In the example topologies, this approach proved to be competitive to the algebraic approach with \code{Kira}.
\code{Kira} is usually faster for smaller problems.
However, when the system of equations becomes huge and large intermediate expressions occur, the reconstruction with \code{FireFly} can become faster and requires much less memory.
The reconstruction approach does not need to compute the large intermediate expressions and can also strictly select only those coefficients which are requested, whereas the algebraic approach can neither circumvent the large intermediate expressions nor perform such a strict selection.

In both approaches, the reduction can be distributed to several computers or several sessions on the same machine, which require less memory.
\code{Kira} allows to perform the back substitution for subsets of master integrals, which is of course limited by the number of master integrals.
The reconstruction approach allows to reconstruct subsets of the coefficients or in principle each coefficient on its own.
The number of coefficients is usually orders of magnitude larger than the number of master integrals.
This also offers the possibility to choose the subsets such that all of them require a sparser shift than the complete set, which can speed up the reconstruction for the individual coefficients significantly.
Even for the straight-on approach for quite simple problems, the reconstruction benefits much more from additional \abbrev{CPU} threads at least up to several hundred of them.
This scaling can even be improved by guessing the required black-box probes in advance.
However, this could lead to unnecessary calculations.

Our implementation within \code{Kira} is strongly limited by the numerical solutions with \code{pyRed}, which require most of the \abbrev{CPU} time in our examples.
Therefore, it is desirable to reduce the number of black-box probes required for the reconstruction by algorithmic improvements in \code{FireFly}.
We hope that racing the Ben-Or/Tiwari algorithm against Newton's algorithm as part of Zippel's algorithm as proposed in \citeres{Kaltofen:2000,Kaltofen:2003} and a dense-sparse-hybrid approach for rational functions as mentioned in \sct{sec:int_rat} achieve this.
Optimizations for \code{pyRed} would also help to widen this bottleneck.
We also noted that performing the forward elimination is usually the most expensive step for \code{pyRed}.
Since the algebraic forward elimination is often cheap compared to the back substitution, it might be worthwhile to perform it algebraically and perform only the back substitution numerically for the reconstruction if one expects to require many black-box probes.

Of course, \code{FireFly} can also be used for hybrid approaches.
For example, \code{Kira 1.2} already offers a multivariate Newton interpolation for (sub-)coefficients over $\mathbb{Z}$ instead of multiplying them algebraically if this seems heuristically useful\,\cite{Maierhofer:2018gpa}.

\section*{Acknowledgments}
We are grateful to Robert V.\ Harlander, Philipp Maierh\"ofer, Mario Prausa, Benjamin Summ, Johann Usovitsch, and Alexander Voigt for technical support, useful discussions, and comments on the manuscript.
This work was supported by the \textit{Deutsche Forschungsgemeinschaft \abbrev{(DFG)}} Collaborative Research Center \href{http://p3h.particle.kit.edu/start}{TRR 257 “Particle Physics Phenomenology after the Higgs Discovery”} funded through project \href{http://gepris.dfg.de/gepris/projekt/396021762}{396021762}.
F.L.\ acknowledges financial support by \abbrev{DFG} through project \href{http://gepris.dfg.de/gepris/projekt/386986591}{386986591}.

The Feynman diagrams in this paper were drawn with Ti\textit{k}Z-Feynman\,\cite{Ellis:2016jkw}.

\clearpage
\appendix

\section{Algorithms}
\label{sec:algorithms}

\subsection*{Maximal Quotient Rational Reconstruction}
Wang's algorithm for the \abbrev{RR}, \alg{alg:euclid}\,\cite{Wang:1981}, only succeeds if the modulus of both numerator $|n|$ and denominator $|d|$ of the rational number are smaller than $\sqrt{m/2}$.
However, the proof of the uniqueness of the result only requires $m \geq 2 |n||d|$\,\cite{von_zur_Gathen}.
This bound is equally distributed to $n$ and $d$ in \alg{alg:euclid}.
In principle, one can adjust the bounds for every number separately, but this requires knowledge of the number before the \abbrev{RR}.

Monagan observed that the guess of the rational number comes together with a huge quotient in the Euclidean algorithm\,\cite{Monagan:2004}.
Based on this observation he suggested the algorithm called Maximal Quotient Rational Reconstruction, which returns the rational number corresponding to the largest quotient encountered.

\begin{algorithm}
  \caption{Maximal Quotient Rational Reconstruction (\abbrev{MQRR})\,\cite{Monagan:2004}.}
  \label{alg:mqrr}
  \algrenewcommand\algorithmicrequire{\textbf{Input:}}
  \algrenewcommand\algorithmicensure{\textbf{Output:}}
  \begin{algorithmic}
    \Require {Integers \code{m} $>$ \code{e} $\geq$ 0 and \code{T} $>$ 0, which is the error tolerance.}
    \Ensure {A rational number \code{n / d} such that \code{n / d} \text{mod} \code{m} = \code{e} and \code{T * |n| * d} $<$ \code{m} or FAIL.}
    \Function{\abbrev{MQRR}}{\code{e}, \code{m}}
      \If{\code{e} $=$ \code{0}}
        \If{\code{m} $>$ \code{T}}
          \State\Return \code{0};
        \Else
          \State Throw an error that the rational reconstruction failed;
        \EndIf
      \EndIf
      \State \code{n} $\gets$ \code{0};\quad \code{d} $\gets$ \code{0};
      \State \code{t} $\gets$ \code{1};\quad \code{old\_t} $\gets$ \code{0};
      \State \code{r} $\gets$ \code{e};\quad \code{old\_r} $\gets$ \code{m};
      \While{\code{r} $\neq$ \code{0} \textbf{ and } \code{old\_r} $>$ \code{T}}
        \State \code{quotient} $\gets$ \code{$\lfloor$ old\_r  / r $\rfloor$};
        \If{\code{quotient} $>$ \code{T}}
          \State \code{n} $\gets$ \code{r};\quad \code{d} $\gets$ \code{t};\quad \code{T} $\gets$ \code{quotient};
        \EndIf
        \State \code{(old\_r , r)} $\gets$ \code{(r, old\_r - quotient * r});
        \State \code{(old\_t, t)} $\gets$ \code{(t, old\_t - quotient * t)};
      \EndWhile
      \If{\code{d} $=$ \code{0} \textbf{ or } \texttt{gcd}(\code{n}, \code{d}) $\neq$ 1}
        \State Throw an error that the rational reconstruction failed;
      \EndIf
      \State\Return \code{n / d};
    \EndFunction
  \end{algorithmic}
\end{algorithm}

It can only be proven that it returns a unique solution if $|n||d| \leq \sqrt{m}/3$.
However, it performs much better in the average case because large quotients from random input are rare.
The parameter $T$ allows to choose a lower bound for the quotients and the algorithm fails if the quotients are smaller.
Of course, it has to increase with an increasing modulus $m$, because a random input will generate higher quotients as well.
Monagan suggested to calculate it by the formula $T = 2^c \lceil \log_2 m \rceil$ and choose $c$ according to the required error tolerance.
We choose $c = 10$ for \code{FireFly}, which roughly corresponds to $1\,\%$ false positive reconstructions.

\subsection*{Solving shifted transposed Vandermonde systems}
The solution of shifted transposed Vandermonde systems defined in \eqn{eq:vandermonde} only requires small changes compared to the solution strategies for usual transposed Vandermonde systems given by \citeres{Zippel:1990,Kaltofen:1988}.
Note that the algorithm presented below is aimed specifically at the shift with one power, i.e.\ the matrix defined in \eqn{eq:vandermonde}.
A solution strategy for general shifts is presented in \citere{Hu:2016}.

Vandermonde matrices of size $M\times M$ are completely determined by $M$ evaluation points $v_1,\dots,v_M$.
Their $M^2$ components are given by integer powers $v_i^j$, $i,j=1,\dots,M$ instead of $j=0,\ldots,M-1$ in the shifted case.
For simplicity, we only consider one variable.
The solution method of such systems is closely related to Lagrange's polynomial interpolation formula.

Let $B_j(z)$ be the polynomial of degree $M$ defined by
\begin{equation}
B_j(z) = \frac{z}{v_j}\prod^{M}_{\substack{m=1\\m\neq j}}\frac{z - v_m}{v_j - v_m} = \sum_{k=1}^M A_{jk} z^k.
\label{eq:van_poly}
\end{equation}
$A_{ij}$ is the matrix defined by the coefficients that arise when the product of \eqn{eq:van_poly} is multiplied out and like terms are collected. The polynomial $B_j(z)$ is specifically designed so that it vanishes at all $v_i$ with $i\neq j$ and has a value of unity at $z=v_j$. Inserting $v_i$ as an argument, one observes
\begin{equation}
B_j(v_i) = \delta_{ij} = \sum_{k=1}^M A_{jk} v_i^k.
\label{eq:van_inv}
\end{equation}
\eqn{eq:van_inv} states that $A_{jk}$ is exactly the inverse of the matrix of components $v_i^k$. Therefore, the solution of the shifted Vandermonde system in \eqn{eq:vandermonde} is just that inverse times the right-hand side,
\begin{equation}
c_{\alpha_i} = \sum_{k=1}^M A_{ik} f(\vec{y}^{\;k}).
\end{equation}
It is left to multiply the monomial terms in \eqn{eq:van_poly} out in order to get the components of $A_{jk}$. By defining a master polynomial $B(z)$ by
\begin{equation}
B(z) \equiv \prod_{m=1}^M (z - v_m) = \sum_{i=0}^M d_i z^i
\end{equation}
one can evaluate its coefficients and then obtain the specific $B_j$ via polynomial divisions by $z-v_j$. Each division is of $O(M)$ and the total procedure is thus of $O(M^2)$. To generalize the solution algorithm to a multivariate system one just needs to redefine $v_i$ as defined in \eqn{eq:vandermonde}. The algorithm for shifted systems is summarized in \alg{alg:vandermonde}.

\begin{algorithm}
\caption{Algorithm to solve shifted Vandermonde systems motivated by \citeres{Zippel:1990,Kaltofen:1988}.}
\label{alg:vandermonde}
\algrenewcommand\algorithmicrequire{\textbf{Input:}}
\algrenewcommand\algorithmicensure{\textbf{Output:}}
\begin{algorithmic}
\Require {An array of probes of a polynomial, an array of the corresponding inserted values and an array of the contributing degrees of the polynomial ordered (co)lexicographically.}
\Ensure {An array with the coefficients of the contributing degrees.}
\Function{solve\_shifted\_transposed\_vandermonde}{\code{probes}, \code{values}, \code{degrees}}
\State \code{num\_eqn} $\gets$ \code{probes} length;$\quad$\code{cis}; $\quad$\code{vis};
\State\code{i} $\gets$ \code{0};
\For{\code{i} $<$ length of \code{values}} \code{//} Calculate $v_i$
\State \code{j} $\gets$ \code{0};
\State \code{vi} $\gets$ \code{1};
\For{\code{j} $<$ length of \code{values[j]}}
\State \code{vi} $\gets$ \code{vi} * \code{values[i][j] ** degrees[i][j]};
\State \code{j} $\gets$ \code{j + 1};
\EndFor
\State \code{vis[i]} = \code{vi};
\State \code{i} $\gets$ \code{i + 1};
\EndFor
\If{\code{num\_eqn} is equal to \code{1}}
\State \code{cis[0]} $\gets$ \code{probes[0] / vi};
\Else
\State \code{dis};
\State \code{dis[num\_eqn - 1] = -vis[0]}; \code{//} Calculate $d_i$
\State \code{i} $\gets$ \code{1};
\For{\code{i} $<$ \code{num\_eqn}}
\State \code{j} $\gets$ \code{num\_eqn - 1 - i};
\For{\code{j} $<$ \code{num\_eqn - 1}}
\State \code{cis[j]} $\gets$ \code{dis[j] - vis[i] * dis[j + 1]};
\State \code{j} $\gets$ \code{j + 1};
\EndFor
\State \code{dis[num\_eqn - 1]} $\gets$ \code{dis[num\_eqn - 1] - vis[0]};
\State \code{i} $\gets$ \code{i + 1};
\EndFor
\EndIf
\State \code{i} $\gets$ \code{0};
\For{\code{i} $<$ \code{num\_eqn}} \code{//} Calculate $A_{jk}$ and $c_{\alpha_i}$
\State \code{t} $\gets$ \code{1}; $\quad$ \code{b} $\gets$ \code{1}; $\quad$\code{s} $\gets$ \code{probes[num\_eqn - 1]}; $\quad$ \code{j} $\gets$ \code{num\_eqn - 1};
\For{\code{j} $>$ \code{0}}
\State \code{b} $\gets$ \code{dis[j] + vis[i] * b};
\State \code{s} $\gets$ \code{s + probes[j - 1] * b};
\State \code{t} $\gets$ \code{vis[i] * t + b};
\EndFor
\State \code{cis[i] = s / t / vis[i]};
\EndFor
\State\Return \code{cis};
\EndFunction
\end{algorithmic}
\end{algorithm}

\section{Parallelization of \code{FireFly}}
\label{sec:ff_internal}
Let us finish this manuscript by a few more words on the details of parallel computing.
Each \code{PolyReconst} and \code{RatReconst} object (cf.\ documentation of \code{FireFly}) possesses two mutexes, namely, \code{mutex\_status} and \code{mutex\_statics}.
The former protects the status of the object, i.e.\ all member variables which can be accessed through getters.
All changes to these variables only occur after \code{mutex\_status} was obtained and all getters obtain \code{mutex\_status} before accessing the variables.
\code{mutex\_statics} protects the static variables which are used by all reconstruction objects, e.g.\ \code{shift} and \code{rand\_zi}, which are private members representing the used shift and a collection of numerical values for different $z_i$ orders.

To interpolate a function, we follow a two step paradigm. In a first step, we provide the reconstruction class with a black-box probe at a given $z_i$ order. This procedure is called a \textit{feed}. If a reconstruction is not already done, the reconstruction class always accepts a feed if it is evaluated for the current prime number. All feeds are stored in a queue for later usage. The interpolation and reconstruction is done independently of the feed procedure. If the member function \code{interpolate} is called, the class tries to reconstruct the black-box function with all stored feeds in the queue and runs until it is done or a probe corresponding to a $z_i$ order is requested, which is not present in the queue. During the interpolation, the object can be fed, but no additional interpolation job is accepted. Thus, the \code{interpolate} function checks whether another interpolation is already running and returns immediately should this be the case.
Otherwise it takes feeds from the queue and proceeds with the interpolation until the queue is empty or the reconstruction is done.
This procedure allows the user to start the interpolation for every feed without concerning whether the interpolation is already running or not.

\clearpage

\IfFileExists{./\jobname_ref.tex}{
  
}{}

\end{document}